\documentclass[10pt]{iopart} 

\usepackage{lipsum,appendix,enumerate,graphicx,caption,algpseudocode,algorithm}
\usepackage[hidelinks]{hyperref}
\usepackage{iopams,soul,dirtytalk}
\usepackage[table,xcdraw]{xcolor}

\expandafter\let\csname equation*\endcsname\relax
\expandafter\let\csname endequation*\endcsname\relax
\usepackage{amsmath,mathrsfs,bm,scalerel}
\usepackage{soul}
\usepackage{etoolbox}
\makeatletter
\def\@mkboth#1#2{}
\newlength\appendixwidth
\preto\appendix{\addtocontents{toc}{\protect\patchl@section}}
\newcommand{\patchl@section}{%
  \settowidth{\appendixwidth}{\textbf{Appendix }}%
  \addtolength{\appendixwidth}{1.5em}%
  \patchcmd{\l@section}{1.5em}{\appendixwidth}{}{\ddt}%
}
\makeatother

\newcommand{\ham}{\mathcal{H}}  
\newcommand{\pf}{\mathcal{Z}}  
\newcommand{\lik}{\mathscr{L}} 
\DeclareMathOperator*{\argmax}{arg\,max}
\newcommand{\av}[1]{\langle #1 \rangle}

\definecolor{mDarkRed}{HTML}{CF0A0A}
\definecolor{mLightBrown}{HTML}{FF7000}

\begin{document}


\title[Statistical models of complex brain networks: a maximum entropy approach]
{Statistical models of complex brain networks: a maximum entropy approach}

\author{Vito Dichio$^{1}$, Fabrizio De Vico Fallani$^{1}$}
\address{$^1$ Sorbonne Universite, Paris Brain Institute - ICM, CNRS, Inria, Inserm, AP-HP, Hopital de la Pitie Salpêtriere, F-75013, Paris, France}

\ead{fabrizio.de-vico-fallani@inria.fr}

\vspace{10pt}
\begin{indented}
\item[]\today
\end{indented}
\begin{abstract}
The brain is a highly complex system. Most of such complexity stems from the intermingled connections between its parts, which give rise to rich dynamics and to the emergence of high-level cognitive functions. Disentangling the underlying network structure is crucial to understand the brain functioning under both healthy and pathological conditions.

Yet, analyzing brain networks is challenging, in part because their structure represents only one possible realization of a generative stochastic process which is in general unknown. Having a formal way to cope with such intrinsic variability is therefore central for the characterization of brain network properties.

Addressing this issue entails the development of appropriate tools mostly adapted from network science and statistics. Here, we focus on a particular class of maximum entropy models for networks, \emph{i.e.} exponential random graph models (ERGMs), as a parsimonious approach to identify the local connection mechanisms behind observed global network structure. Efforts are reviewed on the quest for basic organizational properties of human brain networks, as well as on the identification of predictive biomarkers of neurological diseases such as stroke. 

We conclude with a discussion on how emerging results and tools from statistical graph modeling, associated with forthcoming improvements in experimental data acquisition, could lead to a finer probabilistic description of complex systems in network neuroscience.
\end{abstract}


\tableofcontents

\section{Introduction}
The human brain is a biological system of tremendous complexity. At different scales of neuronal organization, the paradigm of a system \say{made up of a large number of parts that interact in a nonsimple way} \cite{simon1991} turns out to be an apt abstraction. Notably, it suits neurons interacting through synapses at the microscale as well as brain regions' activity coordinating at the macroscale and resulting in the rich spectrum of mind's functional states. 
The study of the brain as a complex system has flourished in the last $20$ years, drawing analogies from the physics of disordered systems, graph theory, dynamical systems and fueled by the modern deluge of data upon nearly all scientific fields \cite{lynn2019}.

By explicitly representing the interactions between the system's components, networks, or graphs, constitute a natural and powerful way to inquire its organizing principles. 
The description of the brain under the lens of network science has lead to a number of fundamental results. Topological network properties, such as node centrality, modular organization and global efficiency, play a fundamental role in the emergence of basic physiological functions, as well as on the apperance of many brain diseases \cite{bullmore2009,bassett2017,stam2014}. 

Beyond purely descriptive analyses, important questions on the network structure include what is the underlying generative process and how local wiring rules result in the observed large-scale properties \cite{bassett2018}.
To this end, network models based on random edge rewiring or nodal preferential attachment rules have been initially investigated and allowed to assess nontrivial properties of brain networks, such as integration and segregation of information or the presence of few nodes with a high number of connections, \emph{i.e.} the so-called hubs \cite{sporns2004b}. 
In addition to purely topological models, brain networks can be more finely modeled by explicitly taking into account the spatial position of the nodes so as to penalize the cost of having long distance connections and minimize the associated metabolic consumption \cite{betzel2016,bullmore2012}.

\begin{figure}
    \centering
    \includegraphics[width=.8\textwidth]{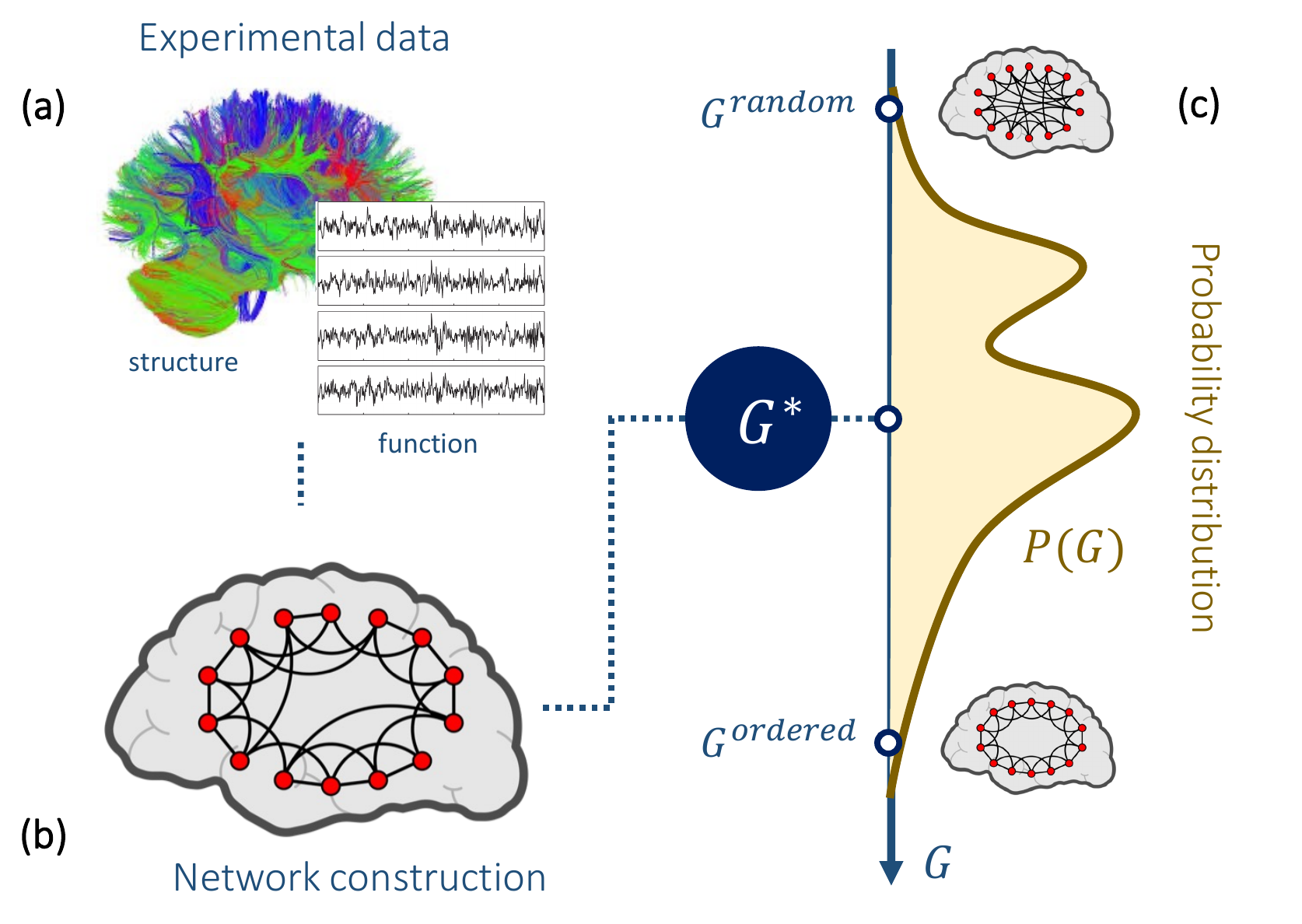}
    \caption{\textbf{Statistical analysis of brain networks inferred from experimental data.} (a) Neuroimaging techniques provide images (\emph{e.g.}, diffusion tensor imaging) and signals (\emph{e.g.}, electroencephalography) informing about the anatomical structure and functional dynamics of the brain, respectively. b) Methods from image and signal processing are then used to build the brain network $G^*$, where nodes correspond to different brain regions and links correspond to structural connections or functional interactions. c) The aim of statistical modeling is to describe $P(G)$, the probability distribution in the space of all possible graphs $G$. It encodes the architectural principles of the observed brain network $G^*$. Typically, the latter has intermediate properties between a totally random $G^{random}$ and a totally ordered $G^{ordered}$ network. }
    \label{brainnet}
\end{figure}

While these models allow to identify putative mechanisms involved in generating the observed brain networks, they implicitly make the assumption that the connections have a physical meaning, \emph{i.e.} they are tangible quantities.
However, it is important to remind that almost all brain networks are currently inferred from experimental data through advanced tools from image and signal processing (\textbf{Fig.\ref{brainnet}a}) \cite{devicofallani2014}. 
In particular, links are statistical estimates of possibly existing anatomical pathways or functional interactions between different brain areas and they are affected by some uncertainty (\textbf{Fig.\ref{brainnet}b}).

Statistical modeling has been introduced in an effort to identify the probability distribution of all the possible network realizations associated with an observed one (\textbf{Fig.\ref{brainnet}c}). Approaches of this kind include generating \textit{a-posteriori} network surrogates preserving some properties (\emph{e.g.}, node degree distribution  \cite{maslov2002}), or \textit{a-priori} reproducing, for example, the underlying modular structure of networks via stochastic block modeling \cite{betzel2018,faskowitz2018}. These approaches have effectively improved the characterization of brain networks with respect to random null-models and provided a statistical framework to estimate confidence intervals. Nevertheless, they remain quite limited in terms of the number of local network properties that can be simultaneously tested and modeled. 

This is a crucial aspect as in general we don't know how many and which are the local connection properties that have generated the global observed structure.
Exponential random graph models (ERGMs) represent an intriguing solution to this limitation, as demonstrated by the recent development in the field.
Broadly speaking, ERGMs belong to the family of maximum entropy (Max.Ent.) models \cite{jaynes1957,cimini2019}. 
Given an observed network $G^*$, the ERGM probability distribution is the one with the largest information entropy that statistically reproduces some key structural properties, or features, of $G^*$. The latter are associated with the model parameters, which can be inferred from the data. 

A fortunate alignment among availability of experimental brain data, development of network theory and of powerful computational tools - mainly borrowed from social sciences \cite{hunter2006} - has allowed researchers to study the brain network structure and dynamics under the lens of the ERGMs.
The goal of the present report is to track down these efforts and present them in a self-contained review with an unified accessible language. The ultimate ambition is to show their usefulness to a broad audience and encourage their exploitation to tackle current open questions in basic and clinical neuroscience. 

The rest of the review is organized as follows. In (Sec.\ref{sBRAINNETS}) we introduce brain networks, briefly describe how to build them from experimental data and which are their main topological properties. In (Sec.\ref{s-ERGM}) we focus on a minimal version of the ERGM, we present both its theoretical and computational aspects, we linger on the interpretation of its parameters and introduce some widely used ERGM graph statistics. In (Sec.\ref{sBRAIN-ERGM}) we review the existing literature of ERGM-based methods for neuroscience. We show how they have been used to characterize brain states, discriminate between different experimental conditions and model dynamic brain networks. In (Sec.\ref{s-INT-ERGM}) the reader will find a discussion of the theoretical subtleties of the ERGMs with a focus on their applications to the brain. Finally, in (Sec.\ref{s-CONC}) we conclude with a perspective on the future of the approach, and point out potentially useful alternatives.

\section{Brain networks in a nutshell}\label{sBRAINNETS}

The study of the brain from a network perspective is nowadays regarded as a vibrant interdisciplinary field of research, referred as network neuroscience \cite{bassett2017}. 
Such efforts have initially focused on quantifying the organizational properties of the nervous system by means of different network metrics \cite{boccaletti2006,lynn2019}. 
For example, it has been shown that brain networks tend to exhibit a pronounced modular structure \emph{i.e.} they are partitioned in densely inter-connected communities linked by sparse inter-community connections \cite{sporns2016}. 
The high clustering together with the low average path length between pairs of nodes has suggested that neuronal interactions are organized in a small-world topology, optimizing the balance between segregation and integration of information \cite{bassett2006,bassett2017b}. 
Moreover, there is mounting evidence on the presence of hubs \cite{gong2009,achard2006} and metabolic constraints of the brain's wiring, with spatially closer regions supporting stronger patterns of connections \cite{betzel2016,rosenbaum2017}.
These brain network properties tend to span different spatial and temporal scales, are in common to several species, correlate with individual behavior and can be significantly affected by neurological disorders \cite{betzel2017,fornito2015}.
In the following, we will focus on brain networks derived from neuroimaging data resulting from the noninvasive recording of large-scale neural information in humans. 

\subsection{Building brain networks from experimental data}\label{ssBRAIN-NETS}

Neuroimaging broadly refers to a large set of techniques that allow to measure noninvasively the brain structure and function. The translation of these measurements into a network hinges on the definition of what nodes and edges are. In general, these definitions depend on experimental constraints as well as on the specific scientific question. Here, we will provide a gentle overview of the main approaches adopted so far in an effort to offer a minimal understanding of the nature of the networks that we will later explore. Should the readers be interested in more detailed technical descriptions, we refer them to some recent books and reviews \cite{devicofallani2014,jones2013,korhonen2021,fornito2016}.

A first type of brain networks aims at representing the structural wiring of the brain. Nodes are defined as different brain sites and edges represent their anatomical interconnections. To this purpose, diffusion tensor imaging (\texttt{DTI}) is a noninvasive technology that allows to infer the diffusion of water molecules through white matter tracts of the brain in the three-dimensional space \cite{basser2000}. The collected data are discretized into non-overlapping gray matter volumes (nodes) and the connection weights (edges) between two nodes are proportional to the anatomical properties of the fiber tracks (\emph{e.g.} number, length). 

The second type of brain networks aims instead at capturing the functional interactions between brain regions (nodes). These regions of interest (ROIs) typically correspond to cytoarchitectural landmarks known a-priori or more directly to sensors used for the measurements \cite{gonzalez2021}. For functional networks, interactions (edges) are defined as statistical similarities between the dynamics taking place in different nodes. Many technologies exist to record nodal dynamics as signals with different temporal and spatial resolutions. 
Among others, electroencephalography (\texttt{EEG}) and functional magnetic resonance imaging (\texttt{fMRI}) allow for noninvasive recordings of electrical activity through sensors placed over the scalp \cite{henry2006} and blood oxygen levels from 3D images of the brain, respectively \cite{raichle1998,salvador2005}. 
Functional interactions are then inferred from the recorded signals using related measures such as Pearson or Spearman correlations, mutual information or Granger causality \cite{devicofallani2014,gonzalez2021}. Thus, the study of functional networks is primarily intended to reveal the brain organizational properties from an information processing point of view. 

At this stage, both structural and functional brain networks are characterized by an adjacency matrix $A$ whose generic entry $a_{ij}$ is a scalar number, typically normalized between $0$ and $1$, representing the magnitude of the connection between nodes $i$ and $j$.
Thresholding procedures can be eventually adopted to filter out irrelevant links and obtain sparse unweighted networks.
More details on why and how to filter information in complex brain networks can be found in \cite{vanwijk2010,devicofallani2017,garrison2015}. 

\subsection{The rationale of a statistical approach}

The rationale for statistical modeling is based on the assumption that the observed system is a realization of a stochastic process, and that by characterizing the properties of this process we can gain insight into the underlying mechanisms governing the system's behavior.
More specifically, the probability of observing each possible realization depends on endogenous system constraints, which can be either postulated or inferred from the data.

This type of statistical uncertainty is different from that associated with exogenous experimental errors - here, the network construction process. Statistical uncertainty is due to the inherent variability of a system and is quantified using statistical measures. Uncertainty due to experimental errors arises from factors such as measurement noise and reflects the limitations of our measurement apparatus. In this review we deal with the former, discussion on the latter can be found elsewhere \cite{devicofallani2014}.

In network neuroscience, a common first approach to assess the statistical significance of a network's properties has been to compare it \textit{a-posteriori} to a null model. The latter is a simplified version of the network that preserves some basic structural features, but removes other non-random ones \cite{vavsa2022}. By comparing a network to its null model, we can determine whether its observed properties are due to chance, or whether they reflect underlying mechanisms or constraints on the system. This comparison can be done using a variety of statistical tests, such as comparing the network's properties to those of a large ensemble of randomized networks, or comparing it to a null model that is specifically designed to test a particular hypothesis.

Such an approach however do not allow to test more specific hypotheses about the mechanisms that govern the network organization, \emph{e.g.} the likelihood that \textit{a-priori} defined structures are at the origin of the observed network.
This kind of questions can be targeted by using instead an inferential approach. This involves (i) constructing a statistical model that describes the relationships between different variables in the system, and (ii) using this model to estimate the parameters that govern the behavior of the system. By comparing the predictions of the model to the observed data, we can determine whether the model accurately captures the underlying mechanisms that drive the system.

Furthermore complex systems typically arise from multiple organizing principles across different scales. Statistical (inferential) frameworks that can simultaneously account for these effects are crucial to understanding the global properties of the system. In network neuroscience, where the focus is on understanding the brain's structure and function, this approach is particularly useful \cite{betzel2017,devicofallani2013,serrano2009}. By using inferential methods, we can identify the combinations of relevant network properties that give rise to the observed large-scale structures and predict how these properties will change under different conditions or perturbations.

Recent endeavors in exponential random graph models (ERGMs) have increasingly attracted the interest of the network science and neuroscience community due to their ability to match the aforementioned desiderata within a coherent and unifying framework.
Hence, it is timely to discuss these emerging developments, and to seek to put them together into a common theoretical ground that can be used to tackle current open questions in modern neuroscience.

\section{Exponential random graph model (ERGM)}\label{s-ERGM}
The interest in the exponential class of probability distributions dates back to the dawn of modern statistics \cite{koopman1936,pitman1936} and was originally motivated by its properties with respect to sufficient statistics, as introduced by R.A. Fisher in the early '20s \cite{fisher1922,fisher1925}. A number of theoretical and computational properties makes them very appealing for the problem of statistical inference \cite{barndorff2014,brown1986} and justifies their ubiquity in all branches of scientific research. In graph theory, the exponential family, here referred to as exponential random graph models (ERGMs), came on stage in the early '80s \cite{holland1981,ove1986,strauss1986} building on the seminal work of J. Besag \cite{besag1974} and further developing since then \cite{wasserman1996,anderson1999}. 
Analogies with well-known methods of statistical mechanics have been recently explored in simple cases in conjunction with the outbreak of network science \cite{bianconi2009,radicchi2020,cimini2019}. 
ERGMs have recently become popular in several scientific fields \emph{e.g.} epidemiology \cite{silk2017}, sociology \cite{handcock2003,robins2007,lusher2013} as well as neuroscience. 
As a consequence an increasing number of publicly available related softwares has appeared, such as the  \texttt{R} language packages collected in the \texttt{statnet} suite \cite{handcock2008,goodreau2008}. Our reference implementation throughout this section is the \texttt{ergm} package \cite{hunter2008}. 

Here, we give the theoretical minimum to familiarize the reader with ERGMs and their estimation methods. 
Notably, we will consider unweighted and undirected graphs as they represent the most common scenario in the current literature. Later in Sec.\ref{sBRAIN-ERGM} we will extend the discussion to more sophisticated cases, when appropriate.

Let $\mathscr{G}$ be a set of all finite graphs of $N$ nodes with no self-loops, containing at most one single edge between two nodes.
Each graph $G\in\mathscr{G}$ can be equivalently represented by a $N\times N$ adjacency matrix $A$ containing Boolean values, \emph{i.e.} $a_{ij}=\{1,0\}$.

At the hearth of the ERGM approach there is the definition of an appropriate probability mass function $P(G)$ over the ensemble $\mathscr{G}$. 
The crucial idea of ERGMs is to encode the \emph{relevant} information about the graph $G$ in a vector $\bm{x}(G)\in\mathbb{R}^r$ of $r$ statistics or metrics, where each element $x_{\alpha}(G)$ measures a different network property of interest.

Specifically, the probability of observing a generic graph $G$ reads as 
\begin{equation}\label{eERGM}
    P(G|\bm{\theta})= \frac{e^{\bm{\theta}\cdot\bm{x}(G)}}{\sum_{\Tilde{G}\in\mathscr{G}}e^{\bm{\theta}\cdot\bm{x}(\Tilde{G})}}\ ,
\end{equation}
where $\bm{\theta}\in\mathbb{R}^r$ are the model parameters weighting the graph statistics and $\cdot$ is the dot product. 
A formal analogy with the canonical Boltzmann distribution is readily recognized by defining the Hamiltonian 
\begin{equation}\label{e-HAM}
    \ham(G) = -\bm{\theta}\cdot\bm{x}(G) = - \sum_{\alpha}\theta_{\alpha} x_{\alpha}(G)\ 
\end{equation}
Hence, the denominator of Eq.\ref{eERGM} is a normalizing constant, which turns out to play the role of partition function $\pf$ and we can rewrite $P(G|\bm{\theta})=\frac{1}{\pf} e^{-\ham(G)}$. 
Given a specific observed network $G^*$, one can infer the model parameters  $\bm{\theta}^*$ so that the expected value of each graph statistics over the ensemble $\mathscr{G}$
\begin{equation}\label{eEXP-VAL}
    \langle \bm{x} \rangle = \sum_{G\in\mathscr{G}} \bm{x}(G) P(G|\bm{\theta}^*) 
\end{equation}
statistically matches the observed value $\simeq \bm{x}(G^*)$. 

Note that (Eq.\ref{eEXP-VAL}) sets the number of model parameters from $2^{\genfrac(){0pt}{2}{N}{2}}$ - the number of possible edges, in the worst case - to a significantly smaller amount determined by the chosen graph statistics.
By consequence, in general a tie-level matching between the networks generated with $P(G|\bm{\theta^*})$ and the observed graph $G^*$ is neither expected nor desired. 

\subsection{Model parameter estimation}\label{ssINF-SIM}

Apart from very simple cases, ERGM parameters are hard to obtain in a closed form. In general, the nature and the number of graph statistics make the denominator of Eq.\ref{eERGM} impossible to compute analytically. 
The evaluation of the partition function $\mathcal{Z}$ is a very well-known problem in many situations, like for example  when trying to infer parameters from Boltzmann-like distributions \cite{nguyen2017,cocco2018}.

In the last 30 years there has been a tremendous theoretical and computational effort in developing methods to address this issue and achieve efficient estimation algorithms. In the following, we present a brief introduction to the key-idea behind many of these methods \cite{geyer1991,geyer1992}.

Given a graph $G^*$ and a set of statistics $\bm{x}(G)$, we can take a Bayesian perspective and argue that our knowledge on the parameters $\bm{\theta}$ is better described by the posterior distribution 
\begin{equation}\label{eBAY-INF}
    P(\bm{\theta}|G^*) = \frac{P(G^*,\bm{\theta})}{P(G^*)} = \frac{P(G^*|\bm{\theta})P(\bm{\theta})}{P(G^*)}\ .
\end{equation}
In the case where no prior information is available for $\bm{\theta}$, we can take $P(\bm{\theta})$ to be uniformly distributed. From Eq.\ref{eBAY-INF} we see that the posterior distribution $P(\bm{\theta}|G^*)$ is directly proportional to the distribution of the data $G^*$ given the parameters $\bm{\theta}$. Accordingly, our best guess on the parameters value is given by the Maximum Likelihood Estimator (MLE)
\begin{equation}\label{eMAXLIK-EST}
    \bm{\theta}^* = \argmax_{\bm{\theta}} \lik(G^*|\bm{\theta})\ , 
\end{equation}
where $\lik(G^*|\bm{\theta}) = \log P(G^*|\bm{\theta})$ is the log-likelihood function. 

The evaluation of (Eq.\ref{eMAXLIK-EST}) is hampered by the computation of the partition function  $\mathcal{Z}$ inside $P(G^*|\bm{\theta})$. As discussed above, this entails the need for non-exact numerical methods.

Rather than maximizing $\lik(G^*|\bm{\theta})$ directly, the idea is to maximize a shifted log-likelihood  $\bar{\mathscr{L}}(G^*|\bm{\theta}) = \mathscr{L}(G^*|\bm{\theta}) - \mathscr{L}(G^*|\bm{\theta_{0}})$, where $\bm{\theta_{0}}$ is an arbitrarily parameter vector. \footnote{In practice $\bm{\theta}_0$ should be close enough to $\bm{\theta}$ to ensure convergence in a reasonable computational time. One popular choice is to take $\bm{\theta}_0$ as the maximum pseudo-likelihood estimation (MPLE) of $\bm{\theta}$ under the additional hypothesis that the edges are mutually independent \cite{hunter2006,desmarais2012}.}

A bit of algebra reveals that the new log-likelihood function can be written as

\begin{equation}
 \bar{\mathscr{L}}(G^*|\bm{\theta}) = (\bm{\theta}-\bm{\theta_0)}\cdot \bm{x}(G^*) - \log \ \langle e^{(\bm{\theta}-\bm{\theta_0})\cdot\bm{x}(G)}\rangle_{\theta_0} 
\end{equation}
where the subscript $\langle\cdot\rangle_{\theta_0}$ indicates the expectation value over the distribution $P(G|\bm{\theta_0})$.
Note that the new log-likelihood $\bar{\mathscr{L}}(G^*|\bm{\theta})$ is maximized by the same $\bm{\theta}^*$ that maximizes $\mathscr{L}(G^*|\bm{\theta})$.

In practice, the evaluation of the partition functions $\mathcal{Z}$ is bypassed and we just need to calculate the expectation $\langle\cdot\rangle_{\theta_0}$, which can be efficiently performed with standard Markov-chain Monte-Carlo (MCMC) approximations. 
Eventually, the whole procedure is repeated until the convergence to some stable solution $\bm{\theta}^*$ is reached (\textbf{Tab. \ref{t-pseudo}}). In the following, we will refer to this method as MCMC-MLE.

\begin{table}
\centering
\begin{algorithmic}
    \State $\bm{\theta}^{(0)} = \bm{\theta}^{PL}$
    \State $ \tau = 0$ 
    \While{(\texttt{conv} = \texttt{FALSE})} 
    \State $\tau \mathrel{+}= 1$
    \State generate $n$ graphs $G_k\sim P(G|\bm{\theta}^{(\tau-1)})$ by MCMC
    \State $\bm{\theta}^{(\tau)} = \argmax_{\bm{\theta}} \Big\{ (\bm{\theta}-\bm{\theta}^{(\tau-1)})\cdot \bm{x}(G^*) - \log( \frac{1}{n}\sum_{k=1}^n e^{(\bm{\theta}-\bm{\theta}^{(\tau-1)})\cdot\bm{x}(G_k)}) \Big\}$
    \If{(convergence criterion)} 
    \State \texttt{conv} = \texttt{TRUE}
    \EndIf
    \EndWhile
\end{algorithmic}
\caption{MCMC-MLE pseudo-code. The initial choice of parameters $\bm{\theta}^{(0)}$ is arbitrary, here the pseudo-likelihood value $\bm{\theta}^{PL}$. Several convergence criteria are possible, from the simplest $\lVert \bm{\theta}^{(\tau)}-\bm{\theta}^{(\tau-1)}\rVert<\epsilon$ for some $\epsilon>0$, up to more refined choices \cite{hunter2006,krivitsky2022}.}\label{t-pseudo}
\end{table}

\subsubsection{ERGMs in practice}
The increasing diffusion of ERGMs has been driven by a concurrent technical effort to develop efficient algorithms for carrying out the inference task. By far the most popular implementation available to date is the \href{https://statnet.org/}{\texttt{statnet}} suite of \texttt{R}-based software packages \cite{handcock2008,krivitsky2022b}. To our knowledge, no other implementation has reached the same level of maturity and extending these techniques outside the \texttt{R} language is an open challenge for the next future. The reference implementation for the vast majority of existing packages is the \texttt{ergm} package \cite{hunter2008}. A number of other packages have been built on it, both for static \cite{caimo2011,wilson2017,schweinberger2018,krivitsky2017b} and temporal networks \cite{krivitsky2014,leifeld2018}.


An illustrative workflow based on the \texttt{ergm} package is implemented in \cite{dichio2023git}, consisting of six fundamental steps. They include: i) creating a network object using the \texttt{network} package from the adjacency matrix and nodal/edge covariates (if any), ii) computing ERGM statistics for the experimental network (optional), iii) specifying and estimating the model, iv) assessing MCMC convergence (optional), v) evaluating the model fit using goodness of fit methods (GoF), and vi) simulating the model using estimated parameters (optional).

The computational time required for the model estimation might vary from a few seconds to several hours, depending on a number of factors including the type and number of graph statistics, the basic properties of the networks (\emph{e.g.} size, sparsity) and several estimation settings that can be possibly adjusted. Because of the intrinsic characteristics of the current neuroimaging technology (Sec.\ref{sBRAINNETS}), brain networks of interest in this review have a relatively small number of nodes (at most hundreds). For such sizes, the ERGM estimation is in general doable. 
For further information regarding the computational aspects and a comprehensive overview of the latest technical enhancements of the \texttt{ergm 4.0} package, we refer to \cite{krivitsky2023,krivitsky2022} and references therein.

\subsection{Interpretation of ERGMs} \label{ss-INTERPR}

A crucial, often overlooked, aspect in ERGM is the interpretation of the parameters $\bm{\theta}$ in Eq.\ref{eERGM}. Let us address the question starting with a simple example \cite{ove1986,park2004,albert2002b}. Consider an ERGM based on a single graph metric $x(G)$ corresponding to the number of edges in the network, \emph{i.e.} $x(G) = \sum_{i<j}a_{ij}$. We first consider the \emph{forward} problem in which there is no observed graph but instead the parameter $\theta$ is given and the goal is to evaluate the expected value for $x(G)$. 

In such a simple case, the mass probability function can be computed exactly:

\begin{equation}\label{eBERN}
    P(G|\theta) = \frac{e^{\theta x(G)}}{\prod_{i<j} \sum_{a_{ij}=0}^1e^{\theta a_{ij}}} =  \frac{e^{\theta x(G)}}{(1+e^{\theta})^{\binom{N}{2}}}.
\end{equation}

Notably, by defining    
\begin{equation}\label{eBERN-PARAM}
    p(\theta) = \frac{e^{\theta}}{1+e^{\theta}}\ 
\end{equation}

we can rewrite $P(G|\theta) = p^{x(G)}(1-p)^{\binom{N}{2}-{x(G)}}$, which is the probability of a Bernoulli graph with $N$ nodes and $x(G)$ links \cite{erdos1960}. The expected value for the number of edges in the graph is then simply $\av{x} = \binom{N}{2}p$. 

In practice, the most common scenario is quite the opposite. One does not know the value of $\theta$ but rather has an observed graph $G^*$ with, say, $M$ edges. In these situations, the goal is to solve the \emph{inverse} problem by inferring $\theta^*$ from $x(G^*) = M$. Taking $p^*= M / \binom{N}{2}$ and inverting Eq.\ref{eBERN-PARAM}:

\begin{equation}\label{e-ber-gra-the}
    \theta^*(p^*) = \log{\frac{p^*}{1-p^*}} 
\end{equation}

so that, consistently, $\av{x} = \binom{N}{2}p = M$. 
Notably,  Eq.\ref{e-ber-gra-the} tells us that if we start from a Bernoulli graph with $p^*=0.5$, then $\theta=0$. This is the maximally random case since each dyad corresponds to the toss of a fair coin. But if we start from a denser ($p^*>0.5$) or sparser ($p^*<0.5$) graph, then $\theta^*>0$ or $\theta^*<0$, respectively. 

In such a simple case we have therefore a complete understanding of both the \emph{forward} and \emph{inverse} problems. In real situations, however, the interpretation of the parameters becomes trickier since ERGMs may include several statistics, often exhibiting some degree of dependence.
As a consequence, there might be interactions among the model parameters that are in general difficult to compute analytically.

Yet, inference is still possible via approximate numerical methods such as the MCMC-MLE. The interpretation of the inferred parameters can be given by quantifying their effect on the likelihood that an edge between two nodes exists or not in the network.
For example, consider a single edge toggle between nodes $i$ and $j$.
Let us call $P(G_{a_{ij}=1,\bm{a}_{\backslash ij}}|\bm{\theta^*})$ the probability of having an edge between the two nodes given the rest of the graph $\bm{a}_{\backslash ij}$, and $P(G_{a_{ij}=0,\bm{a}_{\backslash ij}}|\bm{\theta^*})$ its complementary.

From Eq.\ref{eERGM} one easily finds that 
\begin{equation}\label{e-log-odds}
    \frac{P(G_{a_{ij}=1,\bm{a}_{\backslash ij}}|\bm{\theta^*})}{P(G_{a_{ij}=0,\bm{a}_{\backslash ij}}|\bm{\theta^*})} = \exp\Big(\bm{\theta^*}\cdot \bm{\Delta}^{\bm{x}}_{ij}(G)\Big)\ ,
\end{equation}

where the $\bm{\Delta}^{\bm{x}}_{ij}(G) = \bm{x}(G_{a_{ij}=1,\bm{a}_{\backslash ij}}) - \bm{x}(G_{a_{ij}=0,\bm{a}_{\backslash ij}})$ are the so-called \textit{change statistics}. From Eq. \ref{e-log-odds} is clear that the probability of a link to exist depends on the magnitude and signs of both $\bm{\theta^*}$ and $\bm{\Delta}^{\bm{x}}_{ij}(G)$.

Let us consider a very simple case where the presence of a new link between $i$ and $j$ increases the value of a given graph statistic \emph{i.e.} $\Delta_{ij}^{x}(G)>0$. 
If $\theta^*>0$, then this change statistic favors the probability of this edge to exist in the graph $(a_{ij}=1)$. 
Instead, if $\theta^*<0$ the same change statistic penalizes the existence of the connection, favouring instead ($a_{ij}=0$). 
Overall, the magnitude of this effect is given by the term $\exp{[\theta^*\ \Delta_{ij}^{x}(G)]}$. 
In other words, one can disentangle and quantify the relative influence of each graph statistic on the probability to observe, or not, an edge between two nodes $i$ and $j$. 

Note that the above micro-level interpretation of the parameters is precisely at the core of the MCMC routines used to sample networks from ERGM probability distributions. Markov chains of graphs whose stationary distribution is $P(G|\bm{\theta^*})$ are obtained by sequentially proposing changes on a graph, evaluating the probability of the new configuration by Eq.\ref{e-log-odds} and accepting it according to the Metropolis-Hastings recipe. We refer to recent reviews in the field for a more detailed description of the MCMC methods used in ERGMs \cite{hunter2006}.

\subsection{Graph statistics for ERGMs}\label{ssGRAPH-STATS}
A broad spectrum of graph statistics, or metrics, $x(G)$ have been formulated and included in Eq.\ref{eERGM}.
Here, we present those that have been more frequently adopted in neuroscience.

\begin{figure}
    \centering
    \includegraphics[width=\textwidth]{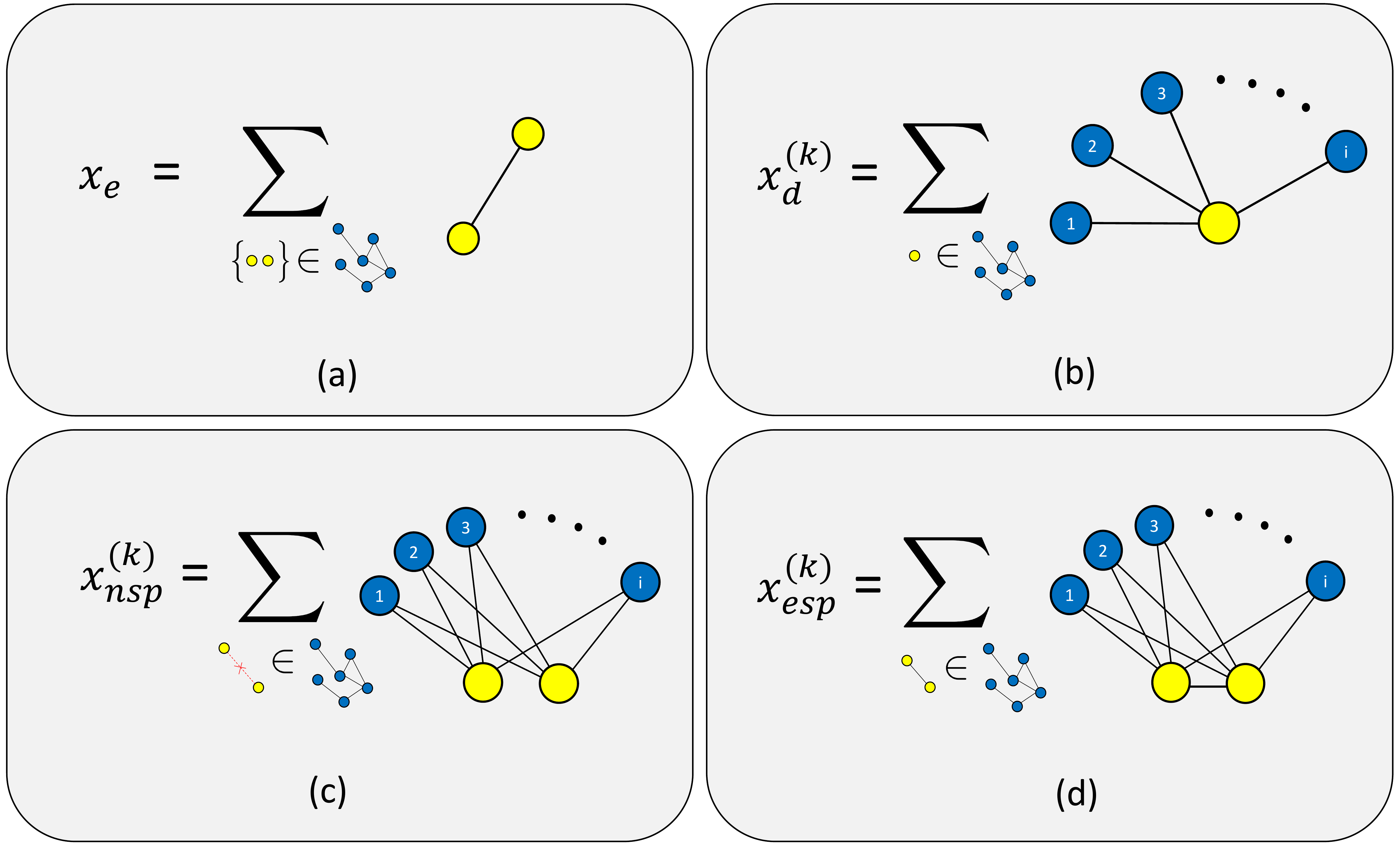}
    \caption{\textbf{Illustration of graph statistics for ERGMs}. a) $x_{e}$ counts the total number of edges in the graph; b) $x_d^{(k)}$  counts the number of nodes with degree $k$; c) $x_{nsp}^{(k)}$ is the number of non-connected dyads sharing $k$ common partners; finally, d) $x_{esp}^{(k)}$ counts the connected dyads sharing exactly $k$ common partners. }
    \label{f-TERMS}
\end{figure}

A first group of graph statistics includes the so-called dyadic independence terms, \emph{i.e.} combinations of only single-dyad terms. 
The most intuitive metric is the \texttt{edges} term which measures the number of edges in the graph (\textbf{Fig. \ref{f-TERMS}a}): 

\begin{equation}\label{e-TERM-EDGES}
    x_{e}(G) = \sum_{i<j} a_{ij} \ .
\end{equation}

A straightforward extension is the so-called \texttt{edge covariate} 
\begin{equation}\label{e-TERM-DYAD-COV}
    x_{dc}(G) = \sum_{i<j} a_{ij} \zeta_{ij}\ .
\end{equation}
where $\zeta_{ij}$ is an attribute associated defined for each dyad, such as the Euclidean distance in a spatial network.

ERGM terms based on nodal attributes also belong to this group. If $\eta_i$ is a categorical property of the nodes, then it is possible to define the so-called \texttt{nodematch} term as 
\begin{equation}\label{eTERM-NM}
    x_{nm}(G) = \sum_{i<j} \delta_{\eta_i\eta_j}a_{ij}\ ,
\end{equation}
where $\delta$ is the Kronecker delta and $x_{nm}$ counts the number of edges whose nodes are labeled by the same categorical attribute. For instance, $\eta_i=\text{R, L}$ could indicate whether a node belongs to the right (R) or left (L) hemisphere.

More in general, real networks are characterized by the presence of complex connectivity structures leading to dependencies between dyads. Therefore, we now turn into presenting graph statistics that involve products of two or more dyadic variables $a_{ij}$. Following the seminal work of O. Frank and D. Strauss on Markov Graphs \cite{ove1986}, we assume that only dyads that share a node can be dependent or, equivalently, that nonincident dyads are conditionally independent.

Under this general definition, several statistics may be introduced based on the count of local connection patterns.
For instance, the presence of hubs, documented in many real-world networks \cite{albert2002a,albert2002b}, can be measured in an ERGM by the \texttt{k-stars} term: 
\begin{equation}\label{e-TERM-ST}
    x_{st}^{(k)}(G) = \frac{1}{k!} \sum_{i_0}\dots\sum_{i_k} a_{i_0i_1}\dots a_{i_0i_k}
\end{equation}
with $k=1,\dots,N-1$. 

The global tendency of networks to form clusters, can be instead measured by the presence of triplets, or triads, of connected nodes. A \texttt{k-triangles} term can be then defined as:
\begin{equation}\label{e-TERM-TR}
    x_{tr}^{(k)}(G) =  \frac{1}{\xi(k)} \sum_{i_0}\dots\sum_{i_{1+k}} a_{i_0i_1}a_{i_0i_{2}}a_{i_1i_2}\dots a_{i_0i_{1+k}}a_{i_1i_{1+k}} 
\end{equation}
where the symmetry factor is $\xi(k)=3!$ for $k=1$ and $\xi(k)=2k!$ for $k=2,\dots,N-2$. The simplest case $x_{tr}^{(1)}(G) = \frac{1}{6} \sum_{i_0 i_1 i_2} a_{i_0i_1}a_{i_0i_{2}}a_{i_1i_2}$ counts the number of triangles in the graph.

\subsubsection{Degeneracy of simple ERGMs.}\label{sss-degen}

Starting from the earliest numerical investigations, it has been observed that for some basic ERGM specifications, MCMC simulations hardly "mix", meaning they do not reach a stationary distribution within a reasonable amount of time \cite{strauss1986,snijders2002}. The lack of mixing hampers the estimation task. This behavior is due to the geometry of ERGM stationary distributions. As it turns out, there exist choices of statistics that result in "degenerate" models, meaning that assign a significant probability only to a few graph configurations, often with a very different structure, such as empty and fully connected graphs. \cite{handcock2003b,rinaldo2009}. A general, thorough discussion of the causes and consequences of degeneracy in the context of ERGMs can be found in  \cite{schweinberger2011}.

For a few simple ERGMs, such as the 2-star model ($\ham=x_e + x_{st}^{(2)}$) or the Strauss clustering model ($\ham=x_e + x_{tr}^{(1)}$), it has been demonstrated through analytic solutions that this phenomenon corresponds to a standard phase separation observed in statistical mechanics between a low- and high-density phase, representing almost empty and full networks, respectively \cite{park2004b,park2005}. For such model specifications, there is no parameters combination that is able to reproduce intermediate densities of edges, two-stars and triangles as observed in real-world networks.

These examples demonstrate that degeneracy issues arise when the model includes unstable sufficient statistics\footnote{According to the definition in \cite{schweinberger2011}, $x(G)$ is stable if there exist $C>0$ and $M_c>0$ such that $U_x\equiv \max_{G\in\mathscr{G}}x(G)\le CM\ \forall M>M_c$, where $M={N \choose 2}$ is the number of degrees of freedom in the network. Note that for instance $U_{x_e} = M$ implies that the edges term is stable while $U_{x_{st}^{(2)}} \sim NM$ implies that the 2-star term is unstable.}. 
Such models show excessive sensitivity in the sense that small state changes can result in extremely large odds ratios eq.(\ref{e-log-odds}) \cite{schweinberger2011}.

The main conclusion of these theoretical investigations is that inference for simple ERGMs is simply not possible, ill-posed. To solve degeneracy issues, a possible solution is to add structure to the model specification by incorporating additional information such as i) edge-specific or node-specific covariates, ii) block structure, iii) temporal structure, iv) multilevel structure, or v) spatial structure. A recent comprehensive discussion of these cases can be found in \cite{schweinberger2020}.

However, the most common approach to ERGM degeneracy issues in neuroscience applications has been so far to include the so-called "curved statistics" in the model specification. In the next section, we will discuss this approach in detail.

\subsubsection{Curved Statistics.}
The inclusion of curved statistics into ERGM model specifications has been found to be effective in alleviating degenerate behaviors \cite{schweinberger2011,schweinberger2020}. 
These statistics are linear combinations of the complete distribution of degree or shared partner statistics, where the coefficients are geometrically weighted. Therefore, these statistics not only enumerate patterns but also include additional non-linear constraints on graph structures through the use of geometric coefficients. \cite{hunter2006, snijders2006}.

For instance, the geometrically weighted degree (\texttt{gwd}) term reads: 

\begin{equation}\label{eTERM-GWD}
    x_{gwd}(G|\tau) = e^{\tau}\sum_{k=1}^{N-1} \Big\{1-\big(1-e^{-\tau}\big)^{k}\Big\}x_{d}^{(k)}(G)\ ,
\end{equation}

where $\tau>0$ is a parameter and $x_{d}^{(k)}(G)$ is the number of nodes in the graph $G$ whose degree is exactly equal to $k$ (\textbf{Fig.\ref{f-TERMS}b}). 
Similarly, the geometrically weighted non-edgewise shared partner (\texttt{gwnsp}) statistic:

\begin{equation}\label{e-TERM-GWNSP}
    x_{gwnsp}(G|\tau) = e^{\tau}\sum_{k=1}^{N-2} \Big\{1-\big(1-e^{-\tau}\big)^{k}\Big\}x_{nsp}^{(k)}(G)\ ,
\end{equation}

where $\tau>0$ and $x_{nsp}^{(k)}(G)$ is the number of non connected dyads having $k$ neighbors in common, (\textbf{Fig.\ref{f-TERMS}c}). Finally, the geometrically weighted edgewise shared partner (\texttt{gwesp}) statistics:

\begin{equation}\label{eTERM-GWESP}
    x_{gwesp}(G|\tau) = e^{\tau}\sum_{k=1}^{N-2} \Big\{1-\big(1-e^{-\tau}\big)^{k}\Big\}x_{esp}^{(k)}(G)\ ,
\end{equation}

where $\tau>0$ and $x_{esp}^{(k)}(G)$ is now the number of connected dyads that share exactly $k$ neighbors (\textbf{Fig.\ref{f-TERMS}d}).

To provide an interpretation of curved statistics and better understand the underlying modeling assumptions, let's examine the case of \texttt{gwesp} and proceed as in (Sec.\ref{ss-INTERPR}). Namely, let $G$ be a graph and consider two connected nodes $i, j$ with $k$ shared partners. Let $G'$ be the same graph as $G$ but with an additional common neighbor between the nodes $i,j$, this implies
    \begin{equation}
    \begin{split}
        x_{esp}^{(k)}(G') &= x_{esp}^{(k)}(G) - 1\\
        x_{esp}^{(k+1)}(G') &= x_{esp}^{(k+1)}(G) + 1 \ ,
    \end{split}
    \end{equation}
    where $x_{esp}^{(k)}(G)$ is the number of connected dyads that share exactly $k$ neighbors. For an ERGM including a \texttt{gwesp} term, we find
    \begin{equation}
    \begin{split}
        \frac{P(G'|\bm{\theta})}{P(G|\bm{\theta})} &\propto \exp\Bigg[\theta_{gwesp} \Big(x_{gwesp}(G'|\tau)-x_{gwesp}(G|\tau)\Big)\Bigg] \\
        &\propto \exp\Big[\theta_{gwesp} (1-e^{-\tau})^{k} \Big] \equiv \rho\ .
    \end{split}
    \end{equation}
    We start by noting that since $\tau>0$, we have $1-e^{-\tau}\in(0,1)$. Moreover:
    \begin{itemize}
        \item[(i)] the scaling factor $\theta_{gwesp}>0$ implies $\rho>1$, hence preference for adding shared partners; $\theta_{gwesp}<0$ on the contrary implies a preference for deleting shared partners. This interpretation is akin to that of all standard ERGM parameters as discussed in (Sec.\ref{ss-INTERPR}).
        \item[(ii)] $\lim_{k\to\infty}\rho = 1$ \emph{i.e.} adding/removing shared partners to pairs of connected nodes with already many of them ($k\gg1$) has little effect on the ratio $P(G')/P(G)$. This implies an effective cutoff for the order $k$ of $x_{esp}^{(k)}$ statistic that is relevant for the system. 
        \item[(iii)] The speed of the geometric decay is controlled by an external 
        parameter $\tau$, small values imply a rapid decay while high values imply a slow decay.
    \end{itemize}  
   Hence, the \texttt{gwesp} statistic is a refined version of the simple sum of triangles. When the parameter $\theta_{gwesp}$ is positive, the general tendency for clustering will not push the system into a fully connected state. 
   In fact, its implementation through the \texttt{gwesp} statistic implies a strong advantage for the addition of shared partners for poorly connected nodes, while almost no effect for nodes that already have many connections. 
   This is indeed a more realistic behavior of a triadic closure, which in the case of brain networks it is expected to be constrained by the inner physical and functional resources \cite{bullmore2012}.

    The same reasoning applies to other curved statistics, such as \texttt{gwd} and \texttt{gwnsp}. These two statistics are alternatively used to test the hypothesis that shortest paths results from the presence of hubs or two-paths.
     The former would affect the node degree distribution, the latter the distribution of non-edgewise shared partners. 
    Thus, via the \texttt{gwd} and \texttt{gwnsp} statistics, the presence of short paths in the network will result in positive $\theta_{gwd}$ and $\theta_{gwnsp}$ values, without the system ending up in a fully connected/empty state.

\section{ERGMs in neuroscience}\label{sBRAIN-ERGM}
\begin{figure}
    \centering
    \includegraphics[width=\columnwidth]{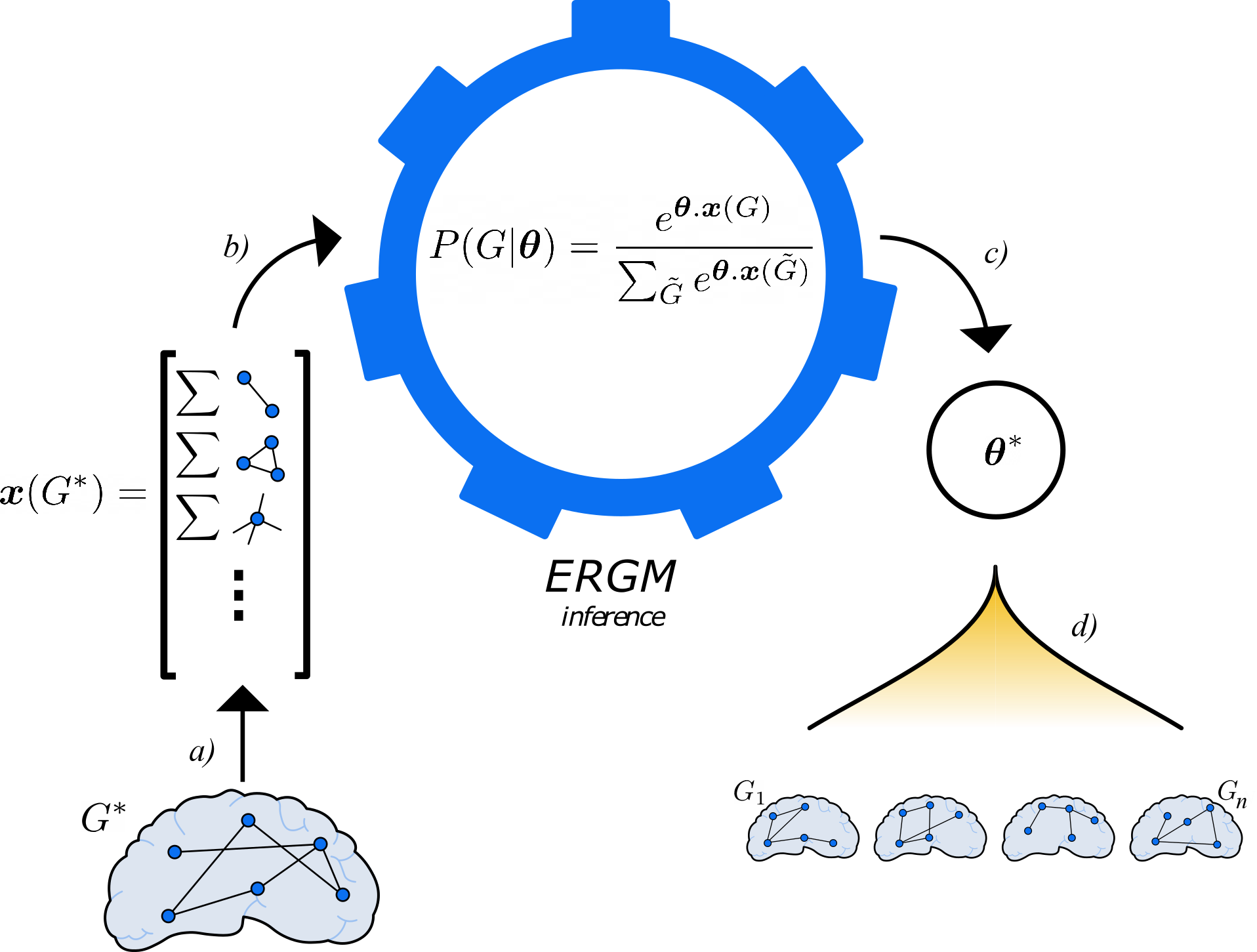}
    \caption{\textbf{ERGMs workflow for brain networks}. a) A graph $G^*$ is given, for instance a structural or functional  brain network. b) The information stored in the graph $G^*$ is condensed in a set of statistics $\bm{x}(\cdot)\in\mathbb{R}^r$, whose choice is up to the modeller. The statistics evaluated on $G^*$ represent the input of the ERGM inference methods. c) The goal of an ERGM is the estimation of the set of parameters $\bm{\theta}^*\in\mathbb{R}^r$ associated with $\bm{x}(\cdot)$. d) The inferred parameters can be used to simulate $n$ new brain networks $G_1,\dots,G_n$ that statistically reproduce the observed network i.e. $\frac{1}{n}\sum_{i=1}^n\bm{x}(G_i)\sim\bm{x}(G^*).$ Goodness of fit (GoF) methods can be used to assess the performance of the estimated model in reproducing other features of the input network \cite{hunter2008b}.}
    \label{f-ERGM}
\end{figure}

ERGMs have started to be exploited as statistical framework to study brain networks since 2010 \cite{vanwijk2010}. The typical workflow of an ERGM analysis in neuroscience is summarized in \textbf{Fig.\ref{f-ERGM}}.
In this section, we present an overview of the most recent results obtained in the last decade. A schematic summary of the existing literature can be found in \textbf{Tab.\ref{t-SUM}}.

While ERGMs are mathematical abstractions and can be applied to any network, most research has so far focused on functional brain connectivity derived from \texttt{fMRI} and \texttt{EEG} data.
In the following, we structure the presentation of the current literature around three main scientific questions: \textit{i)} what are the basic network mechanisms of healthy brain functioning (Sec.\ref{ss-FUN-RES-STA}), \textit{ii)} how to statistically compare brain networks between different states (Sec.\ref{ss-DIS-BRA}), and \textit{iii)} how to explicitly characterize the temporal network evolution following a pathological event (Sec.\ref{ss-INF-DYN}).

\begin{table}
\centering
\renewcommand{\arraystretch}{1.5}
\begin{tabular}{lccl}
\rowcolor[HTML]{e0876a} 
Reference & Data (nodes) & Method/\texttt{Package} \\ \hline
\cite{simpson2011} \emph{Simpson et al. (2011)}  & \texttt{fMRI}($90$) & ERGM / \texttt{ergm} \cite{hunter2008}  \\
\rowcolor[HTML]{fbefcc} 
\cite{simpson2012} \emph{Simpson et al. (2012)}  & \texttt{fMRI}($90$) & ERGM / \texttt{ergm} \cite{hunter2008}  \\
\cite{sinke2016} \emph{Sinke et al. (2016)} & \texttt{DTI}($96$) & BERGM / \texttt{Bergm} \cite{caimo2011}\\
\rowcolor[HTML]{fbefcc} 
\cite{obando2017} \emph{Obando et al. (2017)}   & \texttt{EEG}($56$) & ERGM / \texttt{ergm}  \cite{hunter2008}\\
\cite{stillman2017} \emph{Stillman et al. (2017)}  & \texttt{fMRI}($20$) & cGERGM/ \texttt{gergm} \cite{wilson2017}  \\
\rowcolor[HTML]{fbefcc} 
\cite{dellitalia2018} \emph{Dell'Italia et al. (2018)}  & \texttt{fMRI}(148) & sTERGM / \texttt{stergm} \cite{krivitsky2014} \\
\cite{stillman2019} \emph{Stillman et al. (2019)}  & \texttt{fMRI}($9-25$) & cGERGM/ \texttt{gergm} \cite{wilson2017} \\
\rowcolor[HTML]{fbefcc} 
\cite{lehmann2021} \emph{Lehmann et al. (2021})  & \texttt{fMRI}($90$) & BERGM / \texttt{Bergm} \cite{caimo2011} \\
\cite{obando2022} \emph{Obando et al. (2022)}   & \texttt{fMRI}($81.13^*$) &  TERGM / \texttt{btergm} \cite{leifeld2018}
\end{tabular}
\caption{\textbf{Recent ERGM studies of human brain networks}. We summarize each paper by its reference, the kind of experimental method for the data and the average number of nodes of the resulting networks, ERGM-method employed and its \texttt{R} implementation.\label{t-SUM}}
\end{table}

\subsection{Minimal model of brain networks}\label{ss-FUN-RES-STA}

At the core of many ERGM implementations for neuroscience there is the model first proposed by \emph{Simpson et al.} in two consecutive works \cite{simpson2011,simpson2012}, where the $\ham$-function reads as 

\begin{equation}\label{e-MIN-MOD}
    -\ham(G|\bm{\theta}) = \theta_1 x_{e}(G) + \theta_2 x_{gwesp}(G|\tau) + \theta_3 x_{gwnsp}(G|\nu)
\end{equation}

The \texttt{edges} term $x_{e}(G)$ characterizes the density of the network. This statistic is typically used to ensure that ERGMs also reproduce the actual number of connections in the observed network.
The \texttt{gwesp} term $x_{gwesp}(G|\tau)$ is meant to capture the overall network clustering, a property that is crucial for the segregation of information in the brain \cite{bassett2006,bassett2017b}. 
The \texttt{gwnsp} term $x_{gwnsp}(G|\tau)$ is related to the presence of two paths in the network and can be regarded as a measure of integration of information in the brain \cite{power2013}. 
Hence, Eq.\ref{e-MIN-MOD} defines a parsimonious model of the fundamental properties of brain networks, \emph{i.e.} connectedness, clustering, and global-efficiency. 

In \cite{simpson2011,simpson2012} the authors applied this model to \texttt{fMRI} brain networks from $10$ healthy subjects, fitted separately per each individual.
The same model was also used in \cite{lehmann2021} to characterize fMRI brain networks from $200$ healthy individuals. 
All these works agreed on assigning negative values to the parameters associated with \texttt{edges} and \texttt{gwnsp} statistics and positive values to \texttt{gwesp} statistics, while the decay parameter for the last two cases was fixed to $\tau=\nu=0.75$ \cite{simpson2011}. 
The \texttt{edges} parameter $\theta_1<0$ confirms the tendency of the network to be sparse.
The \texttt{gwnsp} parameter $\theta_3<0$ indicated that any tendency for global integration of information in the brain is unlikely to be supported by shortest paths with length equal to $2$. The \texttt{gwesp} parameter $\theta_2>0$ showed that brain networks are statistically inclined to form clusters, which is a basic hallmark of brain segregation of information.

\begin{figure}
    \centering
    \includegraphics[width=\textwidth,trim={0 1.5cm 0 0},clip]{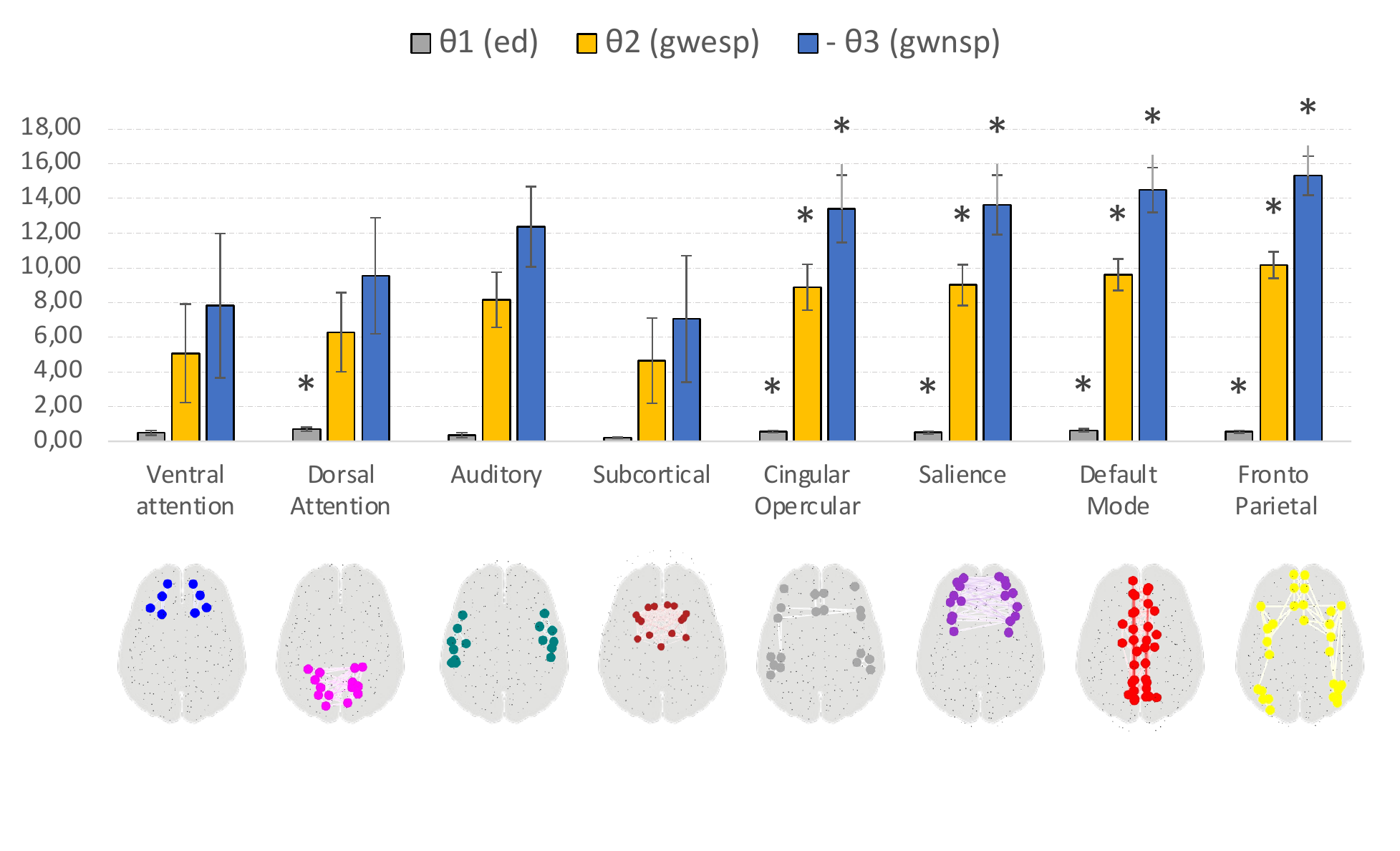}
    \caption{ \textbf{Differences in ERGM parameters across brain subsystems}. Data from \cite{stillman2019}. For each subsystem and parameter, the mean value and st.dev. are reported. Asterisks indicate that the estimation was significantly different from $0$ for all of the $21$ subjects involved in the study. In general, larger networks (cingulo-opercular, salience, default mode, fronto-parietal) yield more robust estimations \emph{i.e.} significant for a larger number of subjects. The corresponding brain subsystems are shown along the x-axis for illustrative purposes, as defined in \cite{gu2015}.}
    \label{f-myfig}
\end{figure}

In subsequent studies, \cite{stillman2017,stillman2019} investigated whether and how the aforementioned whole-brain statistical network properties were also present in specific subsystems associated with basic behavioral functions.
Resting-state fMRI networks were then constructed and fitted separately in $21$ healthy subjects and for $8$ different subsystems, namely auditory, subcortical, dorsal attention, ventral attention, salience, cingulo-opercular task control, fronto-parietal task control, default mode \cite{power2011} (\textbf{Fig.\ref{f-myfig}}). 

To carry out this analysis, the authors introduced the so-called correlation generalized ERGM (cGERGM). This formulation of the model is still based on the ERGM in Eq.\ref{e-MIN-MOD}, with a slightly different geometrical weighting strategy of the graph statistics \cite{wilson2017,stillman2019}. 
However, cGERGM differs from the standard approach in two substantial ways. First, cGERGM was designed to handle full weighted correlation matrices, instead of dealing with unweighted sparse networks.
Secondly, additional geometric information on the nodes were modeled externally to the ERGM by using a beta regression of the group-averaged link weights on the graph statistics \cite{stillman2017}. 
In particular, two kinds of geometric information were included: a \texttt{nodematch} term (Eq.\ref{eTERM-NM}) to measure the tendency of connected nodes to lie in the same hemisphere, and a \texttt{edge covariate} term (Eq.\ref{e-TERM-DYAD-COV}) to measure the total Euclidean distance between all connected nodes. To emphasize the difference from the ERGM parameters, we shall call $\beta_4, \beta_5$ the ones controlling for these effects. 

Across all sub-systems, and almost all subjects, authors found consistent results with parameter values statistically different from $0$ especially for the larger subsystems, \emph{i.e.} cingulo-opercular, salience, default mode, fronto-parietal.
The sign of the $\theta_2>0$ and $\theta_3<0$ parameters associated respectively with \texttt{gwesp} and \texttt{gwnsp} statistics, were in line with those reported for the entire brain network. Instead, the \texttt{edge} statistics was associated with a positive parameter $\theta_1>0$. This result is consistent with the fact that brain subsystems are more densely connected internally than between each other \cite{markov2013,sporns2016}.   

The estimation of the regression parameters associated with the spatial geometry of the brain structural networks turn out to be much more inconsistent across subsystems and subjects. The significantly positive hemispheric parameter ($\beta_4>0$) indicates that the nodes in the  fronto-parietal and default-mode subnetwork are more likely to be connected within hemispheres. 
Instead, there is a general weak tendency of all subnetworks, but the auditory one, to penalize long distant connections ($\beta_5<0$, albeit significant for only between 20\% and 62\% of the individuals).

The overall emerging picture is that of a modular brain network reflecting a significant clustering behavior together with a strong intrahemispheric connectivity. 
Note that the negative \texttt{gwnsp} parameters do not necessarily imply the absence of $2$-paths in the networks. Instead, they indicate that short paths of length $2$ are relatively underrepresented with respect to what could be expected when considering other statistics alone \textit{e.g.,} triangles.

A related question is to what extent integration of information is statistically supported by short paths in the brain network.
From a biological perspective, it would be more plausible that information flows through hubs and not via short paths, although both mechanisms are not mutually exclusive \cite{deco2015}.
To address this question, \cite{obando2017} considered \texttt{EEG} resting-state networks in a group of $108$ healthy subjects. In particular, they compared two different ERGMs, one including \texttt{gwesp} and \texttt{gwnsp} statistics, and the other including \texttt{gwesp} and \texttt{gwd} statistics, that is
\begin{equation}\label{e-OBANDO17}
    -\ham(G|\bm{\theta}) = \theta_2 x_{gwesp}(G|\tau) + \theta_3 x_{gwd}(G|\nu) ;
\end{equation}

while holding fixed the number of edges in the network $x_{e}(G)\equiv x_{e}(G^*)$ and $\tau=\nu=0.75$. 

Results showed that the second ERGM was able to better reproduce brain network properties not included in the model such as modularity, global- and local-efficiency. 
Notably, both $\theta_2$ and $\theta_3$ values were in average positive and significant confirming their statistical relevance for brain networks. 
Taken together, these findings support the hypothesis that while information segregation emerges from triangles, information integration could be actually mediated by hubs rather than shortest paths.

\subsubsection{Group-representative networks.}\label{sss-REP-NET}    
\begin{figure}[ht!]
    \centering
    \includegraphics[width=\textwidth]{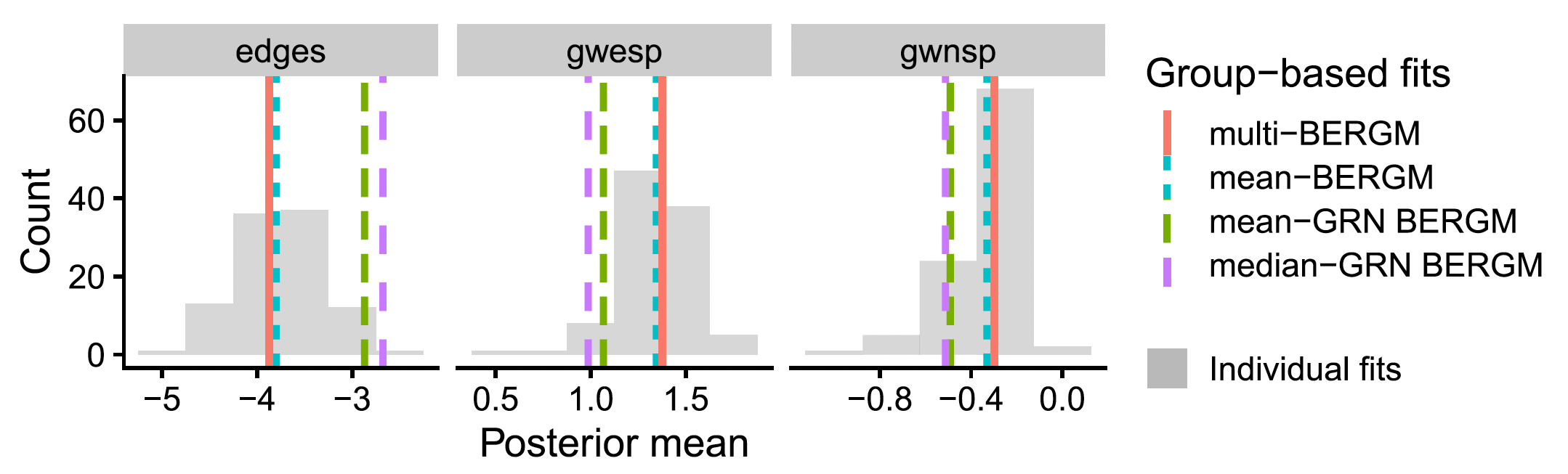}
    \caption{\textbf{Comparison between different GRN construction schemes}. Functional (\texttt{fMRI}) brain networks from $100$ subjects aged $18-33$. Gray histograms show the distributions of parameter estimates from a Bayesian ERGM (BERGM) across individuals. Vertical lines correspond to group-level parameters. In red, the \emph{multi-BERGM} (red line). Blue dotted line for \emph{mean-BERGM} (group-averaged BERGM individual estimates). Green dashed line for a single BERGM estimated on the \emph{mean-GRN}; purple dashed line for a BERGM on the \emph{median-GRN}. Both \emph{multi-BERGM} and \emph{mean-BERGM} show greater accuracy than simpler \emph{mean/median-GRN} in capturing the correct distribution of parameters in the population. Illustration from \cite{lehmann2021}.}
    \label{f-BERGM-LEH}
\end{figure}

Thus far, ERGMs have been used to characterize brain networks obtained from different individuals, separately. The resulting parameters distribution from many subjects can be then used to assess statistical properties at the group-level via standard tools such as hypothesis testing and regression analysis.
Alternatively, assuming that brain networks from different individuals are stochastic realizations of the \emph{same} system, they can be simultaneously modeled so as to obtain a unified group description.
This is the rationale behind the construction of a group representative network (GRN) \emph{i.e.} a network that summarizes the statistical properties of a given ensemble (\textbf{Fig.\ref{f-BERGM-LEH}}).

In network neuroscience, this question has been mainly addressed  by aggregating the connectivity matrices of different individuals to obtain a group-average (\emph{mean-GRN}) or group-median (\emph{median-GRN}) representative network \cite{achard2006,song2009}. 
Alternatively, methods to assess the statistical significance of the pooled  connection values for each pair of nodes can be adopted \cite{zalesky2010}.
Despite being computationally straightforward, the major limitation of these methods is that they treat edges independently and cannot directly inform on complex connection properties involving more than $2$ nodes. An interesting possibility is therefore to incorporate the inter-subject variability directly into the ERGM formulation.

In \cite{simpson2012}, the authors proposed to generate a GRN starting from the ERGMs in (Eq.\ref{e-MIN-MOD}) fitted separately on different subjects. 
Then, a group representative network is simulated via a new ERGM with the $\mathcal{H}$-function 
\begin{equation}\label{e-MIN-MOD-GRN}
    -\ham(G|\bm{\theta}) = \mu_1 x_{e}(G) + \mu_2 x_{gwesp}(G|\tau) + \mu_3 x_{gwnsp}(G|\nu)
\end{equation}
where $\tau=\nu=0.75$ are fixed and the $\bm{\mu}$ are the group-averaged values of the individual model parameters.
This procedure better reproduced several topological properties of the actual brain networks, as compared to standard ERGMs fitted on the group-averaged network. Note that this is a frequentist approach, the estimation of the group level parameters are point-like averages. We refer to this GRN construction scheme as a \emph{mean-ERGM}.

The idea of summarizing individual estimates in a vector of hyper-parameters was pushed further in \cite{lehmann2021}. Accordingly, the model parameter vector for each individual is a realization of a normal probability distribution $\bm{\theta}\sim\mathcal{N}(\bm{\mu},\Sigma_{\bm{\theta}})$. 
The hyper-parameters $(\bm{\mu},\Sigma_{\bm{\theta}})$ are estimated by extending to a multilevel hierarchical setting a Bayesian formulation of the ERGM (BERGM), developed in \cite{caimo2011}.
A GRN can then by obtained by sampling a network from the posterior distribution of parameters. This method is named \emph{multi-BERGM}.

\subsection{Identification of discriminant brain network features}\label{ss-DIS-BRA}

ERGMs can also be used to assess statistical differences between brain networks obtained under different experimental conditions or belonging to different groups of subjects.

In \cite{obando2017} the authors analyzed \texttt{EEG} networks with $56$ nodes in $108$ healthy subjects under two different conditions: eyes closed (EO) and eyes closed (EC) resting-states. 
To this end, they used the same ERGM defined in Eq.\ref{e-OBANDO17}. 
In particular, separate ERGMs were estimated from brain networks extracted from each individual, condition and frequency band $theta$ ($4-7$ Hz), $alpha$ ($8-13$ Hz), $beta$ ($14-29$ Hz), $gamma$ ($30-40$ Hz).
The estimated parameter values were then compared between the two conditions. Results showed that $\theta_2$ values, associated with the \texttt{gwesp} statistics, were significantly higher in EO than EC, for both $alpha$ and $beta$ bands. 
This result is consistent with the local increase of EEG activity widely reported in literature for those bands \cite{barry2007,boytsova2010}. 
Notably, the difference in the $beta$ band could not be detected by a standard network analysis.

ERGM-based analyses have been also adopted to characterize aging. To this end, in \cite{lehmann2021}  the authors compared \texttt{fMRI} brain networks in a group of young (Y: $18-33$ ) and elderly (O: $74-89$) human subjects \cite{lehmann2021}.
Starting from the hierarchical Bayesian ERGM presented in (Sec.\ref{sss-REP-NET}), a further hierarchical layer is added to account for the population partitioned in two subgroups, yielding two different hyper-parameters $\bm{\phi}^{(Y)}=(\bm{\mu}^{(Y)},\Sigma_{\bm{\theta}})$ and $\bm{\phi}^{(O)}=(\bm{\mu}^{(O)},\Sigma_{\bm{\theta}})$ and same covariance structure is assumed for the two groups. 

\begin{figure}[ht!]
    \centering
    \includegraphics[width=\textwidth]{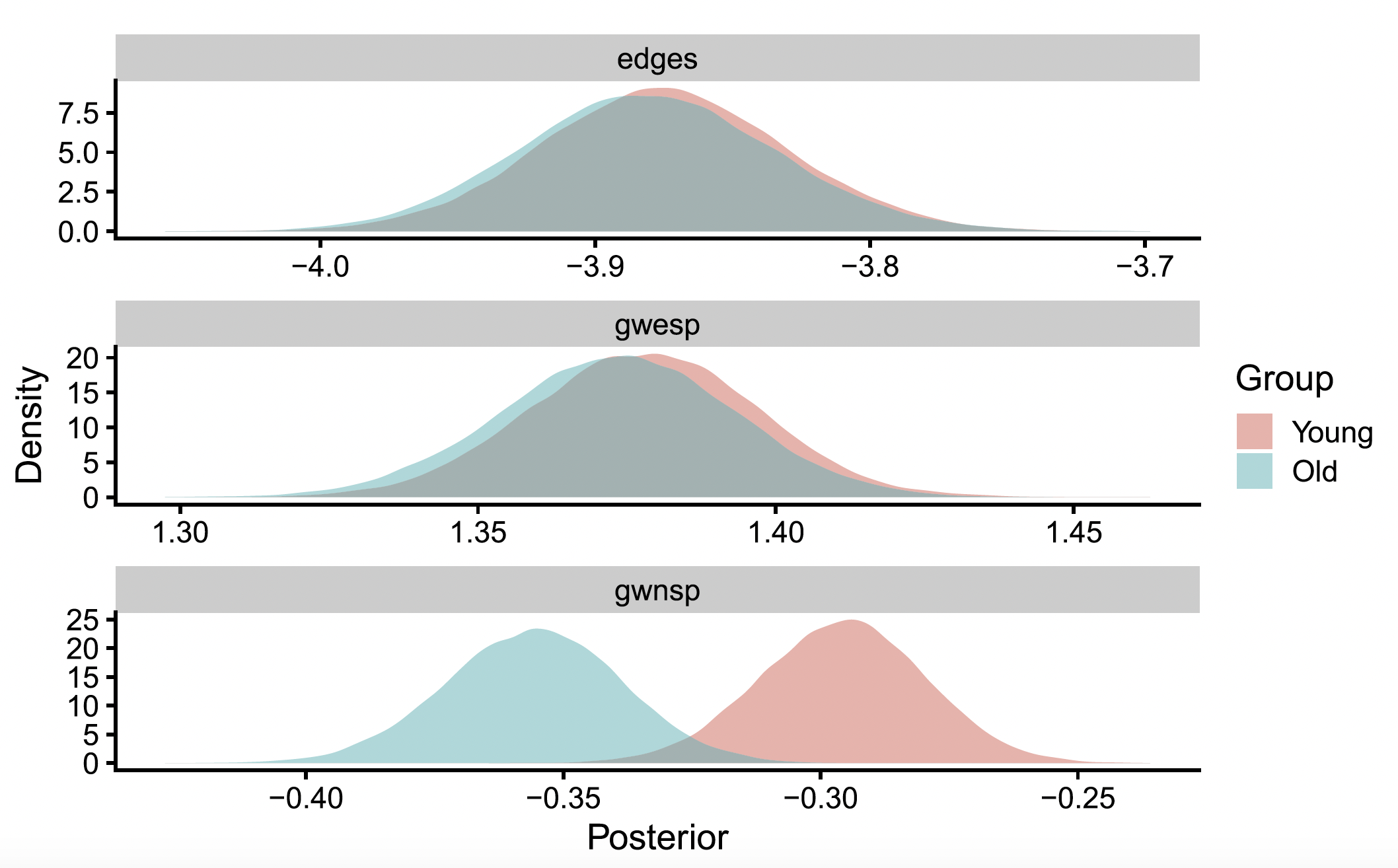}
    \caption{\textbf{Comparison of \emph{multi-BERGM} parameters between young and old individuals}.  Two groups of subjects are considered: young $18-33$ (Y) and old $74-89$ (O). $100$ functional (\texttt{fMRI}) brain networks for each group. The young group is characterized by larger values for the GWNSP parameter while the edges and GWESP posteriors were almost identical between the two. Illustration from \cite{lehmann2021}.}
    \label{f-LEH}
\end{figure}

The results showed that the estimates for the parameter associated with the \texttt{gwnsp} statistics were significantly higher in the young group compared to the old one (\textbf{Fig.\ref{f-LEH}}). These findings indicated that functional brain networks are prone to significant loss of integration of information across the lifespan, consistently with previous findings reporting a decreasing global-efficiency with age \cite{achard2006}. 

\begin{figure}[ht!]
    \centering
    \includegraphics[width=\textwidth]{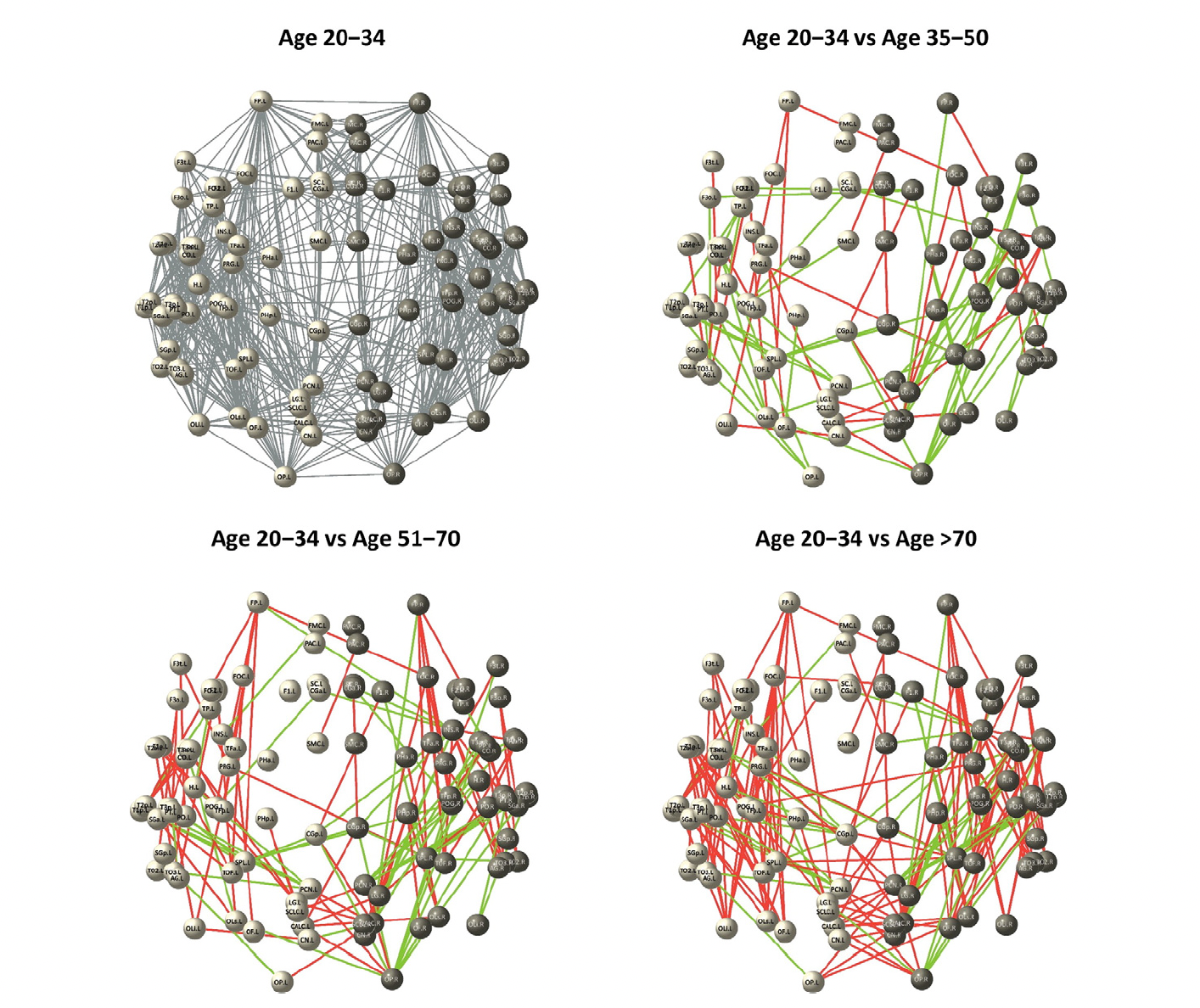}
    \caption{\textbf{Structural brain network changes across lifespan}. Top left: \emph{mean-GRN} for the age range of $20 - 34$. Other networks correspond to brain networks obtained in distinct older groups of individuals. Red lines represent "lost connections" and green lines represent "new connections". Brain network density changes across the lifespan with $+1.6\%$ (age $35 – 50$), $- 4.4\%$ (age $51 – 70$) and $-8.0\%$ (age $>70$). Illustration from \cite{sinke2016}.}
    \label{f-SIN}
\end{figure}

One might then wonder whether similar changes occur in structural brain networks, whose connectivity exhibits much slower changes \cite{gong2009b}.
\emph{Sinke et al.} \cite{sinke2016} studied \texttt{DTI} networks from $382$ healthy subjects and considered $4$ age-groups: $20-34$, $35-50$, $51-70$, $>70$ years (\textbf{Fig. \ref{f-SIN}}). 
Assuming a low variability between the individuals in the same group, authors constructed a \emph{mean-GRN} for each group and used a Bayesian ERGM (BERGM) \cite{caimo2011}. 
 
The  $\mathcal{H}$-function in this case was slightly different from Eq.\ref{e-MIN-MOD} \emph{i.e.} 
\begin{equation}\label{e-MIN-MOD-SIN}
    -\ham(G|\bm{\theta}) = \theta_1 x_{e}(G) + \theta_2 x_{gwesp}(G|\tau) + \theta_3 x_{gwnsp}(G|\nu) + \theta_4 x_{nm}(G)\ ,
\end{equation}
where $x_{nm}(G)$ is a \texttt{nodematch} term (Eq.\ref{eTERM-NM}) measuring in this case the number of connections within the same hemisphere.

The signs of the fitted model parameters ($\theta_1<0$, $\theta_2>0$, $\theta_3<0$) confirmed the general tendencies observed in functional brain networks (Sec.\ref{ss-FUN-RES-STA}).
The parameter $\theta_4$ was instead negative for all groups, suggesting that the hemispheric nodematch is relatively low taking into account all the other included statistics. 
We stress that this does not necessary imply an absolute low intrahemispheric connectivity.

Moreover, differently from previous studies \cite{hagmann2010,otte2015}, only weak changes were found between the parameter values across age-groups. Future studies will elucidate whether this result is actually reflecting a more stable topology of structural brain networks or a result of the group-averaging of the original brain networks.

\subsection{Predicting states in temporal brain networks}\label{ss-INF-DYN}
So far, ERGMs have not explicitly taken into account the fact that brain networks are intrinsically dynamic and their topology can change, fluctuate and evolve over multiple time scales.
Over short-time scales, brain networks rapidly reconfigure during motor or cognitive tasks to adapt to behavioral efforts or loads \cite{devicofallani2008,valencia2008,bassett2011,preti2017}.
Over longer time scales, brain networks can exhibit significant topological losses, such as in aging or in neurodegeneration \cite{wu2013,dautricourt2021}, as well as tentative restorative patterns in recovery after stroke or traumatic injuries \cite{siegel2018}.

In these cases, it is crucial to formulate statistical network models that explicitly include time, so as to capture time-varying connection properties that are predictive of the system behaviour \cite{holme2012,masuda2016}.
One possibility is to endow the ERGM with a time structure and estimate parameters from a time series, or sequence, of observed graphs $G^1,\dots,G^T$. 
One can start by assuming a first-order Markov time dependence, write the probability density function for the entire network sequence as  

\begin{equation}\label{e-TERGM-OB1}
    P(G^{t},\dots,G^{t-1}|\bm{\theta}) = \prod_{t=2}^T P(G^{t}|G^{t-1},\bm{\theta})
\end{equation}
and generalize Eq.\ref{eERGM} in a straightforward fashion:

\begin{equation}\label{eTERGM}
    P(G^{t}|G^{t-1},\bm{\theta})= 
    \frac{e^{\bm{\theta}\cdot\bm{x}(G^t, G^{t-1})}}{\sum_{G^*\in\mathscr{G}}e^{\bm{\theta}\cdot\bm{x}(G^*, G^{t-1})}} \ ,
\end{equation}
where the $\bm{x}$ statistics can be now defined over two consecutive graphs in the time-series. We shall refer to this model as temporal ERGM (TERGM) \cite{hanneke2010}. 

\begin{figure}[ht!]
    \centering
    \includegraphics[width=1\textwidth]{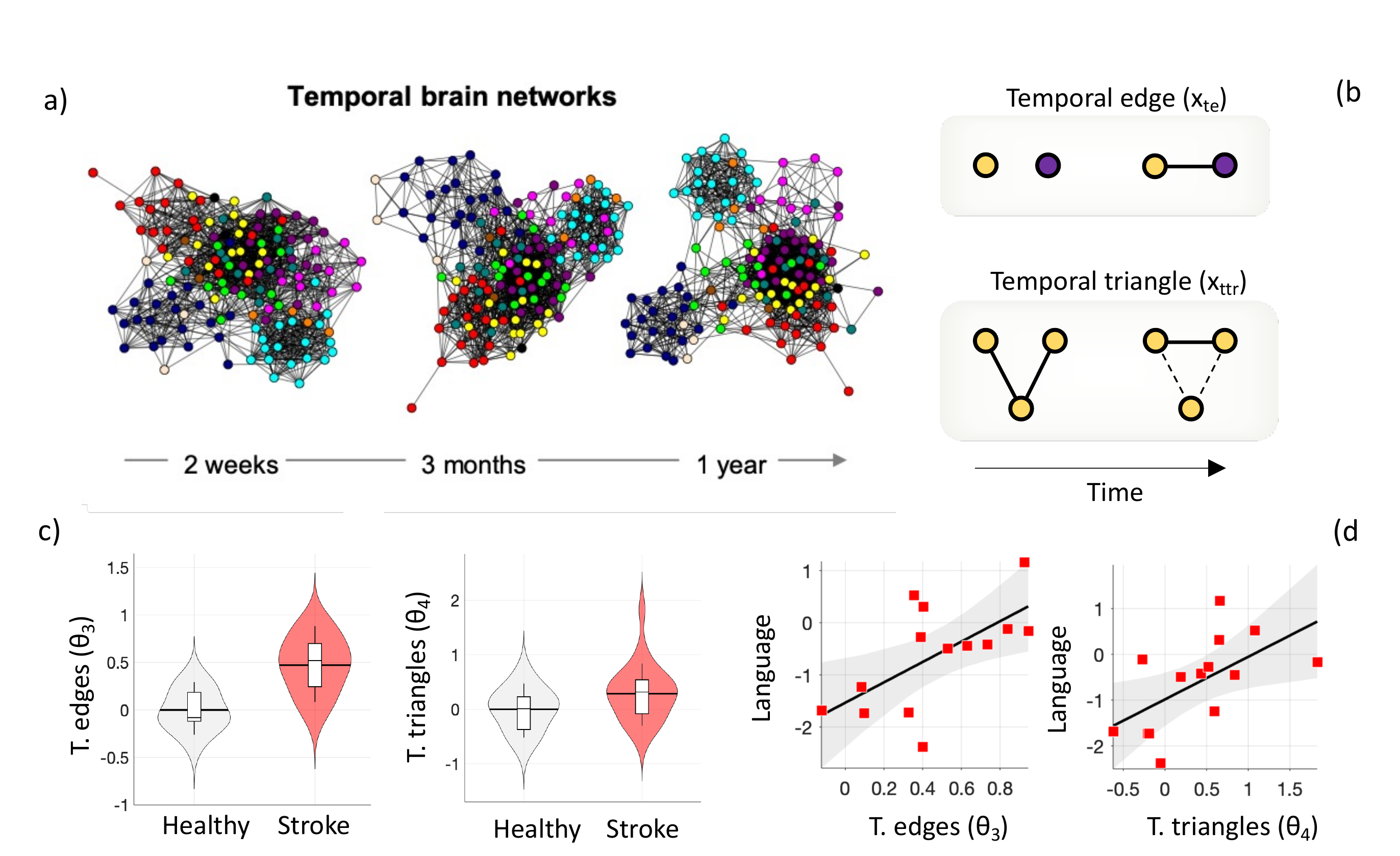}
    \caption{\textbf{TERGM analysis of longitudinal brain networks after a stroke event}. a) Dynamic brain network of a stroke patient 2 weeks, 3 months and 1 year after the stroke event. Different colours emphasize the belonging to different brain systems.  b) Graphical representation of temporal edges $x_{te}$ and temporal triangles $x_{ttr}$. c) Statistical comparison between $\theta_3, \theta_4$ for stroke patients (red shape) and healthy controls (white shape). Violin plots show the distribution of the values, while box-plots represent median and quartiles. Significant increases in the stroke group are reported for both the parameters considered ($p=0.045$). d) Correlation of the TERGM coefficients and future language outcome of stroke patients. Both temporal edges and triangles in the subacute phase (2 weeks, 3 months) are significantly associated with the future language score in the chronic phase (1 year) ($p<0.03$). Illustration adapted from from \cite{obando2022}.}
    \label{f-OBA}
\end{figure}

TERGMs have been recently adopted to model brain network reorganization after stroke \emph{Obando et al.} \cite{obando2022}. The study was carried out on longitudinal \texttt{fMRI} resting-state networks measured in $49$ patients at $2$ weeks, $3$ months and $1$ year after their first ever unilateral stroke event \cite{siegel2016,siegel2018} (\textbf{Fig.\ref{f-OBA}a}). 

The model (Eq.\ref{eTERGM}) was specified with the following temporal $\ham$ function:

\begin{equation}\label{e-TERGM-H}
\begin{split}
    -\ham(G^t| G^{t-1}, \bm{\theta}) =& \ \theta_1 x_{e}(G^t) + \theta_2 x_{stab}(G^t, G^{t-1})\ +
    \\ &+ \theta_3 x_{te}(G^t, G^{t-1}) + \theta_4 x_{ttr}(G^t, G^{t-1})
\end{split}
\end{equation}

Besides the standard \texttt{edges} term (Eq.\ref{e-TERM-EDGES}), the three other temporal statistics were defined as follows.

The \texttt{stability} term $x_{stab}$ simply counts the number of dyads that do not change in the time step $t-1 \rightarrow t$ and it has a fundamental role for the estimation since it sets the pace of allowed changes in the network sequence. 
The \texttt{temporal-edge} $x_{te}$ term counts the number of new inter-hemispheric edges appearing at time $t$, while the \texttt{temporal-triangle} $x_{ttr}$ term counts the number of triangle closures over time within the lesioned hemispheres (\textbf{Fig.\ref{f-OBA}b}). 

These two last statistics are entrusted with the goal of mimicking the brain plasticity process \emph{i.e.} the recovery of between-hemisphere integration and within-hemisphere segregation \cite{siegel2018,grefkes2020}. 
The TERGM analysis resulted in positive estimates ($\theta_3>0$, $\theta_4>0$) and significantly higher in stroke patients compared to healthy controls (\textbf{Fig.\ref{f-OBA}c}). This suggested that both $x_{te}$ and $x_{ttr}$ are fundamental temporal processes of post-stroke brain reorganization. 
Notably, the same conclusions could not be reached when considering a static equivalent formulation of Eq.\ref{e-TERGM-H} \cite{obando2022}.
 
Moreover, the same parameters estimated from the temporal networks in the subacute phase (from $2$ weeks to $3$ months) significantly predicted the future language score ($1$ year) of the patients, suggesting their potential use as clinical biomarkers (\textbf{Fig.\ref{f-OBA}d}).

An alternative approach was adopted in \cite{dellitalia2018} to characterize the recovery from coma after traumatic brain injuries.
Authors evaluated longitudinal \texttt{fMRI} brain networks obtained at $11$, $18$, $25$ and $181$ days after the trauma.
In particular, they used separable temporal ERGMs (sTERGMs) \cite{krivitsky2014}, which model separately and indipendently the tendencies to form and dissolve instances of different graph statistics.

The network at the time-step $t$ is therefore given by the following:
\begin{equation}\label{eSTERGM}
    G^t = G^+ - (G^{t-1} - G^-)
\end{equation}

where $G^{t-1} $ it the graph time $t-1$, while $G^+$ and $G^-$ are the so-called \textit{formation} and \textit{dissolution} graphs.
More in detail, the formation graph is drawn from $\sim P_+(G^{t}|G^{t-1},\bm{\theta}^+)$, where $P_+$ is the same as in Eq.\ref{eTERGM} but with the formation statistics $\bm{x}^+$ and parameters $\bm{\theta}^+$. In this formation process, ties can only be added to $G^{t-1}$.
Analogously, the dissolution graph is obtained by removing edges from $G^{t-1}$ and the graph is drawn from $\sim P_-(G^{t}|G^{t-1},\bm{\theta}^-)$, with dissolution statistics $\bm{x}^-$ and parameters $\bm{\theta}^-$. 
The goal of a sTERMG is to estimate both sets of parameters $\bm{\theta}^+,\bm{\theta}^-$.

The approach in this study was to include in the model a huge number of potentially relevant statistics and let sTERMGs select the most relevant ones. 
Results showed indeed that only a subset of the estimated parameters turned out to be statistically significant. The recovery process was characterized by a tendency to preserve and strengthen connections within pairs of subsystems (\emph{e.g.}, default, frontoparietal, thalamus, visual) and between some of them (eg, default-visual, somatomotor-frontoparietal, ventral-visual). Similarly to the results found in \cite{obando2022}, a significant propensity to close triangles over time also emerged from this study, thus confirming the importance of the local clustering connections in shaping the brain reorganization process from brain damage-induced coma.

\section{On the interpretation of ERGMs}\label{s-INT-ERGM}

Up to now, we have not lingered on the foundation and subtleties of ERGMs, albeit they are crucial to understand and interpret the related results. Due to the number and variety of scientists and practitioners that have developed or used these models, there is sometimes confusion in the scientific literature of ERGMs, if not in the notions at least in the lexicon. This might lead to unsupported claims or reckless interpretations of the results. We therefore aim at clarifying in this section the philosophy of the method and its theoretical underpinnings.

Our discussion lies on three fundamental concepts:\textit{ i)} description, \textit{ii)} prediction, and \textit{iii)} explanation. We briefly pin-down for clarity the corresponding definitions. 
\emph{Description} means a characterization, or a summary, of the state of the system under investigation, based on the choice of suitable variables and observables to be measured. By definition, a description refers to the past or present state of the system. 
\textit{Prediction} is the anticipation (forecast) of the future state of a system.
\emph{Explanation} aims at reaching a causal or mechanistic understanding of why a particular phenomenon or event has occurred, based on a theoretical framework or empirical evidence \cite{hempel1948,hanna1969,shmueli2010}. 


We argue that in general ERGMs as those described in (Eq.\ref{eERGM}) and their derivatives, are certainly descriptive, might be predictive, but they are not explicative.
Note that a method can be tremendously efficient in predicting outcomes of an experiment while at the same time providing no explanation on the underlying physics. Analogously, a description of the phenomena under investigation is radically different from a scientific explanation \cite{hempel1948}.

\subsection{ERGMs and the maximum entropy principle (MEP)}\label{ssERGM-MAXENT}
First, let us explicit the origin of the ERGM formulation from basic information theoretic principles.
ERGMs can be indeed seen as one of the many particular instances belonging to the broad class of Maximum Entropy (Max.Ent.) models \cite{jaynes1957,cimini2019,cover1999}. Similar models have been applied to neural populations \cite{schneidman2006}, antibody diversity \cite{mora2010}, protein sequences \cite{cocco2018}, economics \cite{reddy2020}, population genetics \cite{dichio2021} and many more. Let us then derive the ERGMs from the MEP.\footnote{Note, however, that this is not the way the ERGMs were originally introduced, which instead was based on the  Hammersley-Clifford theorem for Markov graphs \cite{ove1986}, building on a previous work of J. Besag in the context of spatial models of lattice systems \cite{besag1974}.}.

Consider an observed graph $G^*\in\mathscr{G}$ and assume that all the \emph{relevant} information about $G^*$ is encoded in a set of sufficient statistics $\bm{x}(G^*)$. According to the MEP, the most unbiased probability density function $P(G)$ consistent with available knowledge is obtained by maximizing the Shannon information entropy $\mathcal{S}$
\begin{equation}\label{eSE}
    \mathcal{S}[P]=-\sum_{G\in\mathscr{G}}P(G)\log P(G)
\end{equation}
under the constraints
\begin{align}
    \sum_{G\in\mathscr{G}}P(G) &= 1 \label{e-con-norm}\ ,\\
    \sum_{G\in\mathscr{G}}\bm{x}(G)P(G) &= \bm{x}(G^*) \label{e-con-avs}\ :
\end{align}
the first one is just a normalization, the others fix the expected values of the sufficient statistics over the ensamble $\mathscr{G}$ to the correspondent values computed for $G^*$. 

The rationale at the basis of the Max.Ent. recipe is very pragmatic, as already stressed in \cite{jaynes1982}, and dates back in similar forms to Bernoulli and Laplace. 
Let us consider the total number of ways in which the probability distribution of a system can be realized, given that all its possible states are equivalently likely. 
By a simple combinatorial theorem \cite{jaynes1982}, it can be shown that the Max.Ent. distribution is the most likely and also that any other distribution becomes highly atypical as the number of degrees of freedom - in our case, the number of possible edges - becomes large.

In order to solve the constrained maximization problem defined by Eqs.\ref{eSE}-\ref{e-con-avs} we introduce a set of Lagrange multipliers $\bm{\lambda}\in\mathbb{R}^{r+1}$ and maximizing over $P$ the expression:
\begin{equation}
    -\sum_{G\in\mathscr{G}}P(G)\log P(G) + \lambda_0\Big(\sum_{G\in\mathscr{G}}P(G) - 1\Big) + \sum_{\iota=1}^R \lambda_{\iota} \Big(\sum_{G\in\mathscr{G}}x_{\iota}(G)P(G) - x_{\iota}(G^*)\Big)\ ,  
\end{equation}
which gives $P(G|\bm{\lambda}) = e^{\lambda_0-1 + \sum_{\iota=1}^r\lambda_{\iota} x_{\iota}(G)}$. 
By imposing (Eq.\ref{e-con-norm}-\ref{e-con-avs}), we find  $\lambda_0 = 1-\log\pf$ and $\lambda_{\iota}=\theta_{\iota}$ for $\iota=1,\dots,R$, with the $\bm{\theta}$ being \emph{defined} as the choice of the corresponding multipliers which satisfy Eq.\ref{e-con-avs}. The result is precisely Eq.\ref{eERGM}. 

The same argument can be used for a temporal series of graphs $\bm{G} = G^1,\dots,G^T$. If the information is available in the form of graph statistics $\bm{x}(\bm{G^*})$ evaluated on the observed sequence $\bm{G^*}$, the path entropy reads as 
\begin{equation}\label{e-SH-ENT-DYN}
    \mathcal{S}[P] = \sum_{G^1\in\mathscr{G}}\dots\sum_{G^T\in\mathscr{G}} P(\bm{G}) \log P(\bm{G})\ ,
\end{equation}
and the mass probability function becomes

 \begin{equation}\label{eTERGM-MAXENT}
    P(\bm{G}|\bm{\theta})= \frac{e^{\bm{\theta}\cdot\bm{x}(\bm{G})}}{\sum_{\Tilde{G^1}\in\mathscr{G}}\dots\sum_{\Tilde{G^T}\in\mathscr{G}}e^{\bm{\theta}\cdot\bm{x}(\Tilde{\bm{G}})}}\ ,
\end{equation}

which is the more general case of the TERGMs presented before (Eq.\ref{e-TERGM-OB1}-\ref{eTERGM}). This generalized variational principle is sometimes referred as maximum caliber (Max.Cal.) \cite{presse2013}. 

\subsection{Description and prediction: the choice of graph statistics}\label{sssROLE-MOD} 

As discussed in (Sec.\ref{sss-degen}), when formulating ERGMs it is crucial to avoid degeneracy issues which might occur especially with too simplistic model formulations. The possibility of including several potentially relevant effects (model terms) has two advantages. On the one hand, it helps alleviating degeneracy issues, on the other hand it allows to statistically test the joint effect of multiple network properties. Once the model is specified and as long as the estimation process converges properly, an ERGM is always descriptive, the estimated parameters convey information on the relevance of the associated statistics. Whether an ERGM is also predictive is a less trivial question that we address in the following.

A delicate implicit assumption of ERGMs is that the selected statistics $\bm{x}(G^*)$ encode all the \emph{relevant} information of the system, \emph{i.e.} they are \emph{sufficient} statistics. This assumption raised major criticism around Max.Ent. and related methods, see \emph{e.g.} \cite{aurell2016,auletta2017,dichio2021}. 
Indeed, what is deemed to be \emph{relevant} entirely relies on the modeler's hypothesis and is consequently affected by some arbitrariness. 
This is particularly true in complex systems science where, almost by definition, we rarely have control on the hidden degrees of freedom \cite{bassett2018,ladyman2013}. 
It might be that what we aim to measure, or what we are able to measure, simply is not a relevant feature of the system.

In the case of TERGMs, such as those in Eq.\ref{e-TERGM-OB1}-\ref{eTERGM}, there are two more subtleties.
First, the Markov hypothesis in Eq.\ref{e-TERGM-OB1} is in general hard to justify a-priori \cite{onsager1953,auletta2017}. Second, when writing Eq.\ref{eTERGM} we implicitly assume that the selected statistics are relevant throughout the observed time window. This is true if we can assume the stationarity of the process  \cite{gardiner2004}, whose biological plausibility is however disputable \cite{fang2019}. 
Both these two assumptions are further complicated by the fact that in many real applications the measurements of networks are not equally spaced in time \cite{block2018}.

To address these issues, \emph{a-posteriori} model-selection approaches like goodness of fit (GoF) or out-of-sample assessment can be adopted to rule out models that are markedly wrong. 
These methods evaluate the ability of the graphs simulated with a fitted ERGM to reproduce properties of the system other than those explicitly included.
When this happens then the estimated ERGM is not only \textit{descriptive} but also \textit{predictive}.
Thus, when using ERGMs with GOF assessments, it is not only important to use appropriate \emph{relevant} graph statistics but also to have a precise hypothesis on the network features against which to test the predictive performance. 

\subsection{Explanation: ERGMs and Stat.Mech.}\label{ssMAT-EQ}

In the previous sections, we have insisted on the role of the modeler in shaping the mass probability distribution by selecting a set of graph statistics.
This is consistent with the E.T. Jaynes' subjective interpretation of probability \cite{jaynes2003,presse2013}, according to which probabilities are epistemic statements about physical systems. Put differently, they reflect our state of knowledge or, equivalently, of ignorance, about the world.

Such subjective probability is enough for purely inferential purposes. For example, even if Eq.\ref{eERGM} is not the \emph{true} probability density function of the system under investigation, it may nevertheless have good predictive performance (GoF) for a given set of test statistics. Indeed, the ERGM inferential scheme is agnostic about the data-generating process.

This property is somewhat unsatisfactory if the ultimate goal is to explain the physics underlying the observed data. In that case, probabilities should not reflect a subjective state of knowledge but they should have an objective \emph{physical} meaning. 

This is, for instance, the case of statistical mechanics where probabilities have an objective irreducible meaning, they do not depend on the observer. The role of the theory is then to explain how the macroscopic properties of bodies, mainly thermodynamic properties, statistically arise from their microscopic structure \cite{peliti2011}. 
The formal analogy between the ERGM in (Eq.\ref{eERGM}) and the Gibbs-Boltzmann distribution for physical systems at equilibrium has instilled hopes for a similar micro-level objective interpretation of the ERGM probability density function. However, on a closer inspection we immediately realize that this equivalence is problematic. 

These two exponential distributions have in fact very different theoretical underpinnings. Just as an example: at the very foundation of statistical mechanics, and hence of Gibbs-Boltzmann exponential distributions, there is the assumption that the system dynamics obeys a \emph{detailed balance} \cite{boltzmann1970}. In its simplest form, detailed balance means that for any two accessible states $r,s$ the transition rate $W_{r\rightarrow s}$ is equal to its inverse $W_{s\rightarrow r}$. 
This assumption becomes highly disputable when it comes to biological processes. Detailed balance is indeed violated at the molecular level and apparently so also for large-scale brain dynamics even during intrinsic spontaneous activity \cite{lynn2021}. 
Therefore, in general, that between the ERGM and the Gibbs-Boltzmann distributions should not be considered more than a formal analogy, and their respective interpretations should not be confused.

\section{Conclusion and perspectives}\label{s-CONC}
Statistical models are crucial to gain intuition on the connection rules and generative mechanisms of brain networks across multiple time and spatial scales \cite{presigny2022}.
Furthermore, they can be used to improve the detection of the underlying network properties in the presence of statistical noise in the data \cite{efron1986,gfeller2005,karrer2008}.

Here, we have discussed a family of models - exponential random graph models (ERGMs) - which allows  for simultaneously taking into account different local connection properties, disentangle their interactions and quantify their statistical influence on the global observed network. 
The number of studies using ERGMs to model brain connectivity networks has rapidly increased in the last decade.
We have discussed how they have been used to characterize intrinsic brain connectivity, build group representative networks, discriminate different brain states, as well as model temporal patterns, with focused applications to brain diseases. 

In addition, we have placed the ERGM framework in the broader context of maximum entropy models and discussed in detail the subtleties of network analyses based on the Max.Ent. recipe, with particular focus on the meaning of the inferred model parameters. We have discussed the descriptive and predictive power of ERGMs, and warned about its interpretation as an equilibrium \emph{theory} of the brains' biological processes, for which a non-equilibrium picture is more appropriate. 

Despite the increasing number of works using or developing ERGMs to address neuroscience-related questions, the field is still in its infancy and we hope this review will encourage curious physicists, network scientists, and neuroscientists to pursue the research direction we have laid out here. 
In the years to come, we anticipate a boost in the field fueled by an increasing availability of high-quality experimental data, which is crucial to validate the hypothesized mechanisms.
This includes the use of ERGMs to model brain diseases not only from a recovery perspective (\emph{e.g.}, stroke or coma), but also in terms of degeneration (\emph{e.g.} Parkinson, Alzheimer's), as well as to evaluate the efficacy of treatments for psychiatric conditions (\emph{e.g.} Schizofrenia) and developmental disorders  (\emph{e.g.} autism).

On a more methodological note, only a small fraction of the ERGM-like galaxy of statistical models has been employed so far in network neuroscience. 
Among the unexplored alternatives, we spot out the potential use of multilayer ERGMs \cite{krivitsky2020} to capture interactions between different scales of neural organization \cite{presigny2022}. 
As for temporal analyses, a promising family of models beyond the (T)ERGM paradigm is the related stochastic actor oriented model (SAOM), which simulates time-varying network changes through a microscale process defined at the level of the nodes, \emph{i.e.} the actors \cite{snijders2010}. 
Critically, both TERGMS and SAOM only take into account monotonic network changes increasing or decreasing over time. However, modeling brain reorganization might require capturing non-monotonic mechanisms, too. This is particularly true for intrinsic brain functioning and cognitive/motor tasks, which typically exhibit oscillatory time-varying dynamics \cite{preti2017}. In this direction, the development of models that integrate this dynamic component, such as the varying-coefficient ERGM \cite{lee2020}, appear particularly important to overcome such limitations.
A distinct mention is for stochastic block models (SBMs), which allows flexible and scalable analysis of complex networks \cite{holland1983}. In addition to handling community structures, recent SBM extensions also include several types of structures, such as node degree sequences \cite{karrer2011}, hierarchical structures \cite{peixoto2014}, edge weights \cite{peixoto2018}, and triadic closures \cite{peixoto2022}.
SBM applications to brain networks are still in their infancy (see \cite{faskowitz2018} for a recent example), but they surely represent a future promising research avenue. 
Finally, a long-term and ambitious goal would be to enrich statistical models by including higher-order interactions, which might reveal previously unappreciated brain organization principles \cite{battiston2020}.

We believe that combining physics-inspired concepts, biological knowledge, and statistical methods is essential for unraveling the intricacies of the human brain's structure and dynamics. This synergy holds the potential to provide crucial insights into the complexity of the brain.

\section*{Acknowledgements}
\addcontentsline{toc}{section}{\protect\numberline{}Acknowledgements}%
We would like to thank the anonymous reviewers, whose comments helped us improving the initial version of the review. We thank Erik Aurell and Mario Chavez for many insightful discussions and for providing useful remarks on the manuscript. We are also thankful to Thibault Rolland and Remy Ben Messaoud for their contribution in the preparation of the figures. FDVF acknowledges support from the European Research Council (ERC) under the European Union’s Horizon 2020 research and innovation program (Grant Agreement No. 864729)

\section*{References}
\addcontentsline{toc}{section}{\protect\numberline{}References}%
\bibliographystyle{unsrt}
\bibliography{vito,fabrizio}

\end{document}